 \documentclass[preprint]{aastex}

\newbox\grsign \setbox\grsign=\hbox{$>$} \newdimen\grdimen
\grdimen=\ht\grsign
\newbox\simlessbox \newbox\simgreatbox \newbox\simpropbox
\setbox\simgreatbox=\hbox{\raise.5ex\hbox{$>$}\llap
     {\lower.5ex\hbox{$\sim$}}}\ht1=\grdimen\dp1=0pt
\setbox\simlessbox=\hbox{\raise.5ex\hbox{$<$}\llap
     {\lower.5ex\hbox{$\sim$}}}\ht2=\grdimen\dp2=0pt
\setbox\simpropbox=\hbox{\raise.5ex\hbox{$\propto$}\llap
     {\lower.5ex\hbox{$\sim$}}}\ht2=\grdimen\dp2=0pt
\def\simgreat{\mathrel{\copy\simgreatbox}}
\def\simless{\mathrel{\copy\simlessbox}}

\shorttitle{A \emph{Chandra} X-ray Study of NGC 1068. II.}
\shortauthors{Smith and Wilson}

\begin{document}

\title{A Chandra X-ray Study of NGC 1068: II. The Luminous X-ray
Source Population}


\author{David A. Smith and Andrew S. Wilson\altaffilmark{1}}

\affil{Department of Astronomy, University of Maryland, College Park,
MD 20742; dasmith@astro.umd.edu, wilson@astro.umd.edu}

\altaffiltext{1}{Adjunct Astronomer, Space Telescope Science
Institute, 3700 San Martin Drive, Baltimore, MD 21218;
awilson@stsci.edu}


\begin{abstract}

We present an analysis of the compact X-ray source population in the
Seyfert~2 galaxy NGC 1068, imaged with a $\sim 50$~ks \emph{Chandra}
observation.  We find a total of 84 compact sources on the S3 chip, of
which 66 are located within the $25.0$ B-magnitude (arc sec)$^{-2}$
isophote of the galactic disk of NGC 1068.  Spectra have been obtained
for the 21 sources with at least 50 counts, and modeled with both
multi-color disk blackbody and power-law models.  The power-law model
provides the better description of the spectrum for 18 of these
sources.  For fainter sources, the spectral index has been estimated
from the hardness ratio.  Five sources have $0.4$-$8$~keV intrinsic
luminosities greater than $10^{39}$~erg~s$^{-1}$, assuming that their
emission is isotropic and that they are associated with NGC 1068.  We
refer to these sources as Intermediate Luminosity X-ray Objects
(IXOs).  If these five sources are X-ray binaries accreting with
luminosities that are both sub-Eddington and isotropic, then the
implied source masses are $\simgreat 7 \, M_{\odot}$, and so they are
inferred to be black holes.  Most of the spectrally modeled sources
have spectral shapes similar to Galactic black hole candidates.
However, the brightest compact source in NGC 1068 has a spectrum which
is much harder than that found in Galactic black hole candidates and
other IXOs.  The brightest source also shows large amplitude
variability on both short-term and long-term timescales, with the
count rate possibly decreasing by a factor of two in $\sim 2$~ks
during our \emph{Chandra} observation, and the source flux decreasing
by a factor of five between our observation and the grating
observations taken just over nine months later.  The ratio of the
number of sources with luminosities greater than $2.1 \times
10^{38}$~erg~s$^{-1}$ in the $0.4$--$8$~keV band to the rate of
massive ($> 5 \, M_{\odot}$) star formation is the same, to within a
factor of two, for NGC 1068, the Antennae, NGC 5194 (the main galaxy
in M51), and the Circinus galaxy.  This suggests that the rate of
production of X-ray binaries per massive star is approximately the
same for galaxies with currently active star formation, including
``starbursts''.

\end{abstract}

\keywords{accretion, accretion disks --- galaxies: Seyfert ---
galaxies: individual (NGC 1068) --- galaxies: starburst --- X-rays:
binaries --- X-rays: individual (NGC 1068)}

\section{INTRODUCTION}

Observations with the \emph{Einstein Observatory} were the first to
reveal the existence in some nearby galaxies of non-nuclear compact
sources with X-ray luminosities\footnote{Unless noted otherwise, all
luminosities quoted in this paper assume isotropic emission.}
exceeding $10^{39}$~erg~s$^{-1}$, which is well above the Eddington
limit for a $1.4$ solar mass ($M_{\odot}$) neutron star (for a review
of the \emph{Einstein} observations see e.g., Fabbiano 1995).  More
recent observations with \emph{ROSAT} have shown that these
Intermediate Luminosity X-ray Objects (IXOs)\footnote{Here and
elsewhere in this paper, we refer to this class of object as
Intermediate Luminosity X-ray Objects.  Other acronyms for this class
include Ultra-Luminous X-ray sources (ULXs), Super-Luminous X-ray
Sources (SLS), and Super-Eddington X-ray Sources (SES).} are common,
with 9 of the 29 galaxies surveyed by Lira, Lawrence, \& Johnson
(2000) hosting at least one source (see also Colbert \& Mushotzky 1999
and Roberts \& Warwick 2000 for comparable studies with \emph{ROSAT}).

If these sources are powered by accretion, and are emitting at close
to their Eddington limit, then they would contain black holes with
masses up to several hundred $M_{\odot}$ or higher (e.g., Zezas,
Georgantopoulos, \& Ward 1999; Kaaret et al. 2001; Matsumoto et
al. 2001).  Black holes of this mass cannot arise from the collapse of
a single, non-zero metallicity star, but may have instead originated
from an earlier generation of zero metallicity (population III) stars
(e.g., Madau \& Rees 2001).  Alternatively, a massive black hole may
have formed in the center of a globular cluster.  If the cluster then
merges with the galaxy disk, the black hole may then accrete gas in
dense molecular clouds \citep{mh02}. \citet{ebi01} and \citet{pm02}
have suggested black hole formation in young star clusters by rapid
collapse.

If the X-ray emission is beamed, then the required black hole mass may
be reduced to values found in Galactic black hole candidates.
\citet{rey97} and \citet{kor02} have suggested that some of the X-ray
emission from IXOs may be relativistically beamed, and that IXOs may
be objects similar to the Galactic superluminal sources such as
GRS~1915$+$105 and GRO~J1655$-$40, but with their jets pointed towards
the observer.  Alternatively, IXOs may represent a brief evolutionary
phase of close separation X-ray binaries, in which the neutron star or
black hole accretes mass from an early type star at a highly
super-Eddington rate without the system evolving into a common
envelope \citep{kin01}.  This model invokes a thick accretion disk
with a central funnel, which may radiate a total luminosity in excess
of the Eddington limit (e.g., Abramowicz et al. 1980), although the
stability of such structures is questionable.

Other possibilities include supernovae exploding into dense
circumstellar material (e.g., Franco et al. 1993; Plewa 1995).  Such
objects can reach luminosities of up to $10^{41}$~erg~s$^{-1}$ (e.g.,
SN~1988Z: Fabian \& Terlevich 1996; SN~1995N: Fox et al. 2000),
similar to that observed in the most luminous IXOs.  However,
supernovae cannot show the kind of X-ray variability noticed in some
IXOs (see below).  More recently, \citet{beg02} has argued that an
accretion disk around a solar mass black hole could exhibit strong
density inhomogeneities, which would allow the escaping flux to exceed
the Eddington limit by a factor of $10$--$100$, thus explaining the
high luminosities observed without beaming.

Some of the recent work on the X-ray spectra of IXOs has concentrated
on the contribution from an optically thick accretion disk around a
black hole (e.g., Makishima et al. 2000).  Although many of the IXO
spectra can be modeled by a multi-color disk blackbody (MCD) continuum,
the highest (i.e., innermost) disk temperature derived from the
observations is usually larger than that expected for an accretion
disk around a black hole accreting at, or below the Eddington limit.
One possible resolution of this problem is that the black hole is
spinning in the same direction as the disk, and hence the inner radius
is closer to the black hole, and the temperature higher, than for a
non-rotating black hole of the same mass (e.g., Zhang, Cui, \& Chen
1997; Makishima et al. 2000).  Alternatively, the high temperatures
may indicate the presence of a slim (i.e., advection dominated, but
optically thick) disk, in which substantial X-radiation is emitted
from within the last stable circular orbit around a Schwarzschild
black hole (Watarai et al. 2000; Watarai, Mizuno, \& Mineshige 2001;
Mizuno, Kubota, \& Makishima 2001).

More recent observations with \emph{Chandra} have identified IXOs that
cannot be modeled with an MCD spectrum, but require instead a hard
power-law continuum of photon index $\Gamma \approx 1$--$2$ (e.g.,
Kubota et al. 2001; Strickland et al. 2001).  While some models of
supernovae exploding into a dense circumstellar medium do predict hard
power-law continua with $\Gamma \approx 1.6$--$2$ \citep{fra93,ple95},
there is evidence of a spectral transition between a high (with a soft
X-ray spectrum) and a low (with a hard X-ray spectrum) intensity state
in two IXOs in IC~342 \citep{kub01}, and in one IXO in the dwarf
galaxy Holmberg IX \citep{lap01}.  These observations lend further
weight to the X-ray binary interpretation of IXOs, in which the hard
power-law continuum is interpreted as evidence for either a low (hard)
state, as commonly seen in Galactic black hole candidates, or highly
Comptonized, optically thick disk emission, which often characterizes
the very high state of Galactic black hole candidates (e.g., Kubota,
Done, \& Makishima 2003).  Conversely, one source in M51 exhibited
hard ($\Gamma = 1.2$), luminous ($L(0.5-8 \, \rm keV) = 2.7 \times
10^{39}$ erg s$^{-1}$) emission in June 2000, but one year later
showed a very soft ($\Gamma > 5.1$), lower luminosity ($L(0.5-8 \, \rm
keV) = 5.6 \times 10^{38}$ erg s$^{-1}$) spectrum (Terashima \& Wilson
2002a).  Such behaviour is reminiscent of soft X-ray transients.

Variations in the X-ray emission of several IXOs have been observed
both during a single observation (e.g., Okada et al. 1998; Zezas et
al. 1999), and between observations which span several years (e.g., La
Parola et al. 2001).  However, the most spectacular evidence for X-ray
periodicity in an IXO has been found by \citet{bau01} for a bright
compact source in the Circinus galaxy (CG~X-1 or CXOU
J141312.3$-$652013; see also Smith \& Wilson 2001).  The period is
$27.0 \pm 0.7$~ks, and \citet{bau01} argue against the notion that
this source is a foreground Galactic AM~Her type system based
primarily on the low surface density of Galactic X-ray sources in this
direction ($l=311^{\circ}, b=-3.\!^{\circ}8$).  Another IXO in IC~342
is possibly periodic, with a period of 31 or 41~hours, in agreement
with that expected for a semi-detached binary consisting of a
black hole and a main-sequence star of tens of solar masses
\citep{sug01}.

Constraints on the evolution and nature of IXOs can be obtained by
comparing the X-ray luminosity functions (XLFs) of the X-ray source
populations in various types of galaxies and in various evolutionary
stages.  For example, the slopes of XLFs for starburst galaxies (e.g,
M82, ``the Antennae'') are flatter than those found in early-type
galaxies (e.g., NGC 1553, NGC 4967), but are similar to that of the
high-mass X-ray binary population in our own Galaxy (Kilgard et
al. 2002; Grimm, Gilfanov, \& Sunyaev 2002; Zezas \& Fabbiano 2002).
IXOs tend to be associated with regions of active star formation in
spirals, but there are also IXOs in elliptical galaxies \citep{cp02}.
This strong association with star formation suggests that the
brightest X-ray sources in spiral and starburst galaxies are likely to
be young, short-lived sources, e.g., X-ray binaries with O and B type
companions, or supernova remnants \citep{kil02}.  In early-type
galaxies (ellipticals and lenticulars), the XLF tends to break at the
Eddington luminosity of a $1.4 \, M_{\odot}$ neutron star (e.g.,
Sarazin, Irwin, \& Bregman 2000; Blanton, Sarazin, \& Irwin 2001), but
this seems not to be the case in active star-forming regions.

The galaxy NGC 1068 is not only the most luminous ($\simeq 10^{11} \,
L_{\odot}$) nearby Seyfert~2, but also one of the most luminous (also
$\simeq 10^{11} \, L_{\odot}$) starbursts in the local universe (e.g.,
Telesco \& Decher 1988).  It has long been suggested (e.g., Weedman
1983) that the two phenomena are related, but the exact process
remains elusive.  The starburst is circumnuclear, in the galaxy disk
and on a scale of $\sim 2$~kpc.  It is thus a very different
phenomenon to many other regions of star formation in the universe,
such as ``grand-design'' late-type spirals (e.g., M51) and star
formation induced by galaxy mergers (e.g., the Antennae).  For this
reason, we felt it worthwhile to investigate the compact X-ray source
population in a galaxy with such a luminous, circumnuclear starburst.
In this paper, we adopt a distance of $14.4$~Mpc to NGC 1068, so
$1^{\prime\prime} = 70$~pc (e.g., Bland-Hawthorn et al. 1997).

\section{OBSERVATIONS AND DATA REDUCTION}
\label{sec-obs}

\subsection{Observations and Initial Analysis}

NGC 1068 has been observed on four occasions with the \emph{Chandra
X-ray Observatory} \citep{wei01} in direct imaging mode with the
Advanced CCD Imaging Spectrometer (ACIS; Garmire 1997) at the focal
plane of the High-Resolution Mirror Assembly \citep{van97}.  Results
from an analysis of the nuclear and extended X-ray emission associated
with NGC 1068 are presented elsewhere (Young, Wilson, \& Shopbell 2001,
hereafter Paper~I).  A full account of the observations is given in
Paper~I.

We concern ourselves here with a study of the discrete X-ray source
population in NGC 1068, imaged within the $8.\!^{\prime}4 \times
8.\!^{\prime}4$ ($35.3 \times 35.3$~kpc) field of view of the S3 chip.
The data reduction and analysis were done using the Chandra
Interactive Analysis of Observations (CIAO) software version 2.2.1
(released on December 13, 2001) and CALDB 2.12 (released on February
14, 2002), and the reprocessed (on March 10, 2001 using CALDB 2.3 and
version R4CU5UPD14.4 of the processing software) event files.  New
level 2 event files were created, applying the latest telescope
geometry and detector gain corrections, and including the same
\emph{ASCA} grades, bit status, and time filters as in the existing
level 2 event files.  Periods of high and low background (i.e., flares
or data dropouts due to telemetry saturation) were excluded from the
data.  This was achieved by creating a light curve over the full
energy range for the whole S3 chip, excluding the brightest sources of
X-ray emission, and removing events $\pm 3\sigma$ from the mean count
rate.  This procedure gives an effective exposure time of 47.1~ks
(corrected for the dead-time in the detector).  We have checked
whether the decline in low energy quantum efficiency of the S3 chip
affects our results by using the ACISABS spectral model on a few
sources, but found the spectra to be insignificantly different,
presumably because our observations were taken early in the mission
(2000 Feb. 21).

\subsection{Source Detection} 

An image from the 3.2s frame time observation of NGC 1068 (obsid 344),
on the scale of the galactic disk, is shown in Fig.~\ref{fig1}.  There
is a considerable amount of diffuse X-ray emission in this image,
which extends at least $60^{\prime\prime}$ (4.2 kpc) to the northeast,
$50^{\prime\prime}$ (3.5 kpc) to the southwest, $20^{\prime\prime}$
(1.4 kpc) to the northwest, and $30^{\prime\prime}$ (2.1 kpc) to the
southeast of the nucleus.  The dominant large-scale structures are
``spiral arms'' that curve to a lower position angle (P.A.) with
increasing galactocentric distance \citep{you01}.  Many compact
sources of X-ray emission associated with NGC 1068 are also seen in
this image.  The luminous X-ray sources appear to be concentrated in
the starburst region, but there are also many sources located further
out from the center of the galaxy.

Images were extracted from the reprocessed level 2 events file in soft
(0.4--1.5~keV), hard (1.5--5.0~keV), and full (0.4--5.0~keV) energy
bands.  We have used the CIAO program \emph{wavdetect} to search the
images in the three energy bands for discrete sources of X-ray
emission.  This program uses a scalable Marr, or ``Mexican Hat,''
wavelet function to parameterize the shape and extent of each source,
and generally performs better in crowded fields than the standard
``sliding cell'' algorithm which has traditionally been used to detect
sources in X-ray images \citep{fre02}.  We have analyzed the images
using wavelet scales in the range $1$~pixel ($0.\!^{\prime\prime}492$)
to $16$~pixels ($7.\!^{\prime\prime}87$), separated by a factor of
$\sqrt2$ (i.e., $1$, $\sqrt2$, $2$, $2\sqrt2$, $4$, $4\sqrt2$, $8$,
$8\sqrt2$, and $16$ pixels).  These scales correspond to the radius of
the Mexican Hat function, and are representative of the range of
source sizes seen in each of the images.  The wavelet source detection
threshold was set to $10^{-6}$, which will give approximately one
false source for the whole S3 chip.\footnote{See \S11.2 of the CIAO
Detect Manual publicly available at \\
http://asc.harvard.edu/ciao/download/doc/detect\_html\_manual/Manual.html.}
The total number of sources detected by \emph{wavdetect} in the soft,
hard, and full energy band images was 115, 67, and 138, respectively.
Each of these sources was examined carefully by eye, and only those 84
sources which appear compact to the eye are included in the source
list (Table~\ref{tbl-1}).

\subsection{Spectra}
\label{sec-spectra}

Spectra were extracted for those 21 sources listed in
Table~\ref{tbl-1} with at least $50$ counts (after background
subtraction).  We have chosen to ignore 4 sources close to, or in the
``prong'' of bright, diffuse X-ray emission extending in P.A. $\simeq
30^{\circ}$ near the Seyfert nucleus, since their spectra are highly
uncertain, due to contamination by the diffuse emission, and their
signal-to-noise ratio is $<7$.  The shape and size of the extraction
region was chosen to include as much of the source flux as possible,
while minimizing the contribution from background and nearby sources.
Circular regions with diameters in the range $1.\!^{\prime\prime}8$ to
$6.\!^{\prime\prime}0$ sufficed for sources within $\sim 2^{\prime}$
of the telescope mirror axis.  At off-axis angles greater than this,
there is considerable broadening and distortion of the point spread
function and so elliptical regions with major axis diameters in the
range $3.\!^{\prime\prime}7$ to $16.\!^{\prime\prime}6$ and major to
minor axis ratios typically $1.2$--$1.6$ were used.  Background
spectra were usually accumulated from circular or elliptical annuli
surrounding the source.  In the case of sources close to the nucleus,
where the diffuse extended emission is strongest, background spectra
were occasionally taken from circular regions adjacent to the source.
Sources which lie within any given background region were excluded
when calculating the background.  We created a response matrix for
each spectrum using the CIAO programs \emph{mkrmf} and \emph{mkarf}
and the latest calibration files (version N0002, released in August
2001).  Response matrices are available for $32 \times 32$ pixel
regions on the S3 chip.  Differences between response matrices for
adjacent regions are very small, and so, for any given source, we used
the response matrix for the region of the chip where most of the
source photons are located.  In the case of sources which straddle a
node boundary, we extracted a response matrix from each node, and then
created an average response matrix using weights proportional to the
background subtracted counts from each node.  We have ignored the
effects due to the secular decline in the ACIS quantum efficiency
since the observations of NGC 1068 took place early in the mission,
and the effects on the derived values of the column density are
negligible (less than $10$\%) compared to the statistical error.
Prior to performing the spectral analysis with {\sc XSPEC} version
11.1.0 \citep{arn96}, we rebinned the source spectra so that there
were at least 15 counts per bin (for sources with $\geq 100$~counts)
or at least 10 counts per bin (for sources with $<100$ counts), thus
allowing use of the $\chi^{2}$ statistic.  The results of the spectral
analysis are given in Table~\ref{tbl-2}.

\subsection{Timing Analysis}
\label{sec-timing}

Light curves were extracted in the $0.4$--$5$~keV band for each of the
21 non-nuclear sources with sufficient counts for spectral analysis
(see Table~\ref{tbl-2}).  The low count rates for most of the sources
preclude an analysis of the light curves based on the $\chi^{2}$
statistic, which requires the data to be binned.  To assess the
probability of variability, we have, instead, compared the
distribution of event arrival times for each source with that of a
large source-free region of the S3 chip (10557 events), using the
Kolmogorov-Smirnov ($K$--$S$) test (see e.g., Press et al. 1992,
p. 623).

\section{RESULTS}

\subsection{Detected Sources}
\label{sec-detect}

We find a total of 84 compact sources on chip S3 (Table~\ref{tbl-1}),
of which 66 are located within the 25.0 B-magnitude (arc sec)$^{-2}$
isophote of the galactic disk of NGC 1068 (de Vaucouleurs et
al. 1991).  Fig.~\ref{fig2} shows the locations of the 84 compact
sources superposed on the Digitized Sky Survey image.  As can be seen
in this figure, approximately half of the sources are projected onto
the bright inner disk region of NGC 1068.  In order to quantify the
expected number of background sources in the field, we have performed
an identical analysis on an ACIS-S observation of 3C~47 (obsid 2129),
which has a similar exposure time ($41.6$~ks) to our observation of
NGC 1068.  We have considered only regions $\ge 1^{\prime}$ from 3C~47
as there is bright extended X-ray emission surrounding this source.
Regions along the readout trail were also excluded from our analysis.
We find a mean surface density of $0.37$ sources (arc min)$^{-2}$ in
this observation, which is almost equal to the mean surface density of
sources [$\approx 0.45$ sources (arc min)$^{-2}$] outside the $25.0$
B-magnitude (arc sec)$^{-2}$ isophote of the galactic disk of NGC
1068.  This suggests that most of the latter sources are background
objects unrelated to NGC 1068.  From the mean surface density of
sources in the 3C~47 field, we would expect, in our \emph{Chandra}
observation, approximately $12$ background sources projected onto the
galactic disk of NGC 1068, whereas we detect 66 sources.  This number
of background sources is a conservative estimate, since the emission
from such sources must be partially absorbed by gas in the galactic
disk of NGC 1068.

The expected number of background sources can also be estimated from
the density of sources detected in the deep imaging sky surveys with
\emph{Chandra} (see e.g., Cowie et al. 2002 and references therein).
In order to determine the limiting flux level during our
\emph{Chandra} observation, we need to compute the correlation of the
Mexican Hat function with a two-dimensional Gaussian (representing the
point spread function) using the procedure described in Appendix~A of
\citet{fre02}.  The result of this correlation is then compared with
the value expected for a given significance and background level (see
Appendix~B of Freeman et al. 2002).  For sources beyond the $25.0$
B-magnitude (arc sec)$^{-2}$ isophote of the galactic disk, the point
spread function is large, and we adopt a Gaussian of width $\sigma =
3.0$~pixels, which is appropriate for a source $5^{\prime}$ off-axis
and an energy of $4.5$~keV.  We find the detection threshold to be
$\sim 13.5$~counts for our significance and background level.  This
number of counts corresponds to a unabsorbed flux of $1.8 \times
10^{-15}$~erg~cm$^{-2}$~s$^{-1}$ in the 0.4--5~keV band or $2.1
\times 10^{-15}$~erg~cm$^{-2}$~s$^{-1}$ in the 2--8~keV band, assuming
a power-law of photon index $\Gamma = 1.2$ and the Galactic column
density towards NGC 1068 of $N_{\rm H} (\rm Gal) = 3.53 \times
10^{20}$~cm$^{-2}$ \citep{dl90}.  At this flux level, we would expect
a surface density of $0.38$ sources (arc min)$^{-2}$ from the results
of \emph{Chandra} blank field observations in the 2--8~keV band
\citep{cow02}.  While this number is similar to the surface density of
sources outside the $25.0$ B-magnitude (arc sec)$^{-2}$ isophote in
the NGC 1068 observation, the comparison should be considered as only
approximate, since the energy band (2--8~keV) used by \citet{cow02} is
different from the one used in our analysis (0.4--5~keV).  

\subsection{Spectral Analysis}
\label{sec-spectral}

The source spectra were initially modeled in the $0.4$--$8$~keV band
with two alternative descriptions of the continuum: (i) a multi-color
disk (MCD) blackbody spectrum,\footnote{The MCD model in {\sc XSPEC}
returns the color temperature of the inner accretion disk $kT_{\rm
in}$ and an inner disk radius $r_{\rm in} = K^{1/2} \, D \,
(\cos\theta)^{-1/2}$, where $K$ is the normalization of the MCD
spectrum, $D$ is the distance to the source, and $\theta$ is the
(unknown) inclination of the disk (face on corresponds to $\theta =
0^{\circ}$); the luminosity emitted from the disk into our line of
sight varies as $r_{\rm in}^{2} \cos \theta$.  The true inner disk
radius is $R_{\rm in} = \xi \, \kappa^{2} \, r_{\rm in}$, where $\xi =
0.412$ \citep{kub98} and $\kappa = 1.7$ (e.g., Shimura \& Takahara
1995) is the ratio of the color temperature to the effective
temperature.}  which represents the emission expected from an
optically thick accretion disk \citep{mit84,mak86} and (ii) a
power-law spectrum (photon spectral index $\Gamma$).  We have included
in the model absorption by both the Galactic column density towards
NGC 1068 of $N_{\rm H} (\rm Gal) = 3.53 \times 10^{20}$ cm$^{-2}$ and
a column density $N_{\rm H}$ intrinsic to the source; the atomic
cross-sections and abundances for the absorption columns were taken
from \citet{mm83} and \citet{ag89}, respectively.  The results are
given in Table~\ref{tbl-2}.  The MCD and power-law models both provide
a reasonable description of the data for each of the sources, with
neither model being rejected at $\geq 99\%$ confidence using a
$\chi^{2}$ test.  Most of the source spectra have color temperatures
and inner disk radii in the range $kT_{\rm in} \approx 0.5$--$2$~keV
and $R_{\rm in} \approx (10$--$200) \times (\cos\theta)^{-1/2}$~km,
respectively, and intrinsic column densities up to $N_{\rm H} \approx
6.0 \times 10^{21}$~cm$^{-2}$ (MCD models), or photon spectral indices
in the range $\Gamma \approx 1$--$3$ and intrinsic column densities up
to $N_{\rm H} \approx 9.0 \times 10^{21}$~cm$^{-2}$ (power-law
models).  We note that the best-fit intrinsic column densities tend to
be larger for the power-law model than for the MCD model.  It is also
notable that 18 of the 21 sources in Table~\ref{tbl-2} are better
modeled (i.e., have a smaller $\chi^{2}/\nu$, where $\nu$ is the
number of degrees of freedom) by a power-law than an MCD model.
Moreover, when the total set of spectra is considered, the MCD model
can be rejected at $\geq 99\%$ confidence using a $\chi^{2}$ test.

The two most luminous objects have contrasting spectral shapes
(Fig.~\ref{fig3}): the brightest source (J024238.9$-$000055) has a
very hard spectrum ($\Gamma = 0.85^{+0.12}_{-0.11}$) whereas the other
source (J024237.9$-$000118) has a much softer spectrum ($\Gamma =
4.7^{+1.3}_{-0.9}$).  A significant improvement (at $>90$\% confidence
using the $F$ test for two additional free parameters) in the fit to
the spectrum of J024240.7$-$000055 is obtained when we include a
narrow Gaussian emission line.  The line energy and normalization are
$1.031^{+0.028}_{-0.030}$~keV and $(1.02^{+0.79}_{-0.61}) \times
10^{-6}$ photons cm$^{-2}$ s$^{-1}$, respectively.  The equivalent
width is $280^{+270}_{-180}$~eV with respect to the observed
(absorbed) power-law continuum, and the emission line energy is
consistent with that of Ne {\sc x} Ly$\alpha$.  Finally, there is an
absorption feature at $\sim 1.6$~keV in the spectrum of
J024238.9$-$000055 (see Fig.~\ref{fig3}), which may result from an
artifact in the current release of the gain
tables.\footnote{http://cxc.harvard.edu/cal/Links/Acis/acis/Cal\_projects/index.html}

In order to study the population statistically, we have adopted the
following approach.  For the sources whose spectra we have analyzed,
the luminosities in the $0.4$--$8$~keV band were estimated from the
results of fitting a power-law continuum model (see
Table~\ref{tbl-2}).  For the remaining sources, which were spectrally
unmodeled, we have assumed a power-law continuum model modified by (i)
the Galactic column density plus (ii) an intrinsic column density of
$N_{\rm H} = 1.28 \times 10^{21}$~cm$^{-2}$.  The latter column
density is the average (e.g., Bevington \& Robinson 1992, p. 58) of
the spectrally measured intrinsic column densities (power-law model;
Table~\ref{tbl-2}) weighted by $1/\sigma^{2}$, where $\sigma$ is the
larger of the positive and negative errors on the intrinsic column
density.  We believe that these two values of column density are
broadly representative of the range of column densities through which
the X-ray sources in NGC 1068 are seen.  From the hardness ratio $HR =
C(1.5$--$5.0~{\rm keV})/C(0.4$--$1.5~{\rm keV})$, where $C$ represents
the number of counts in the given band, we then determined the
intrinsic photon index of each spectrally unmodeled source.  These
photon indices were then used to estimate the luminosities in the
$0.4$--$8$~keV band.  In cases in which only an upper (lower) limit to
the hardness ratio was determined from the data, we used this upper
(lower) limit to determine the photon index.  Figs.~\ref{fig4}a and b
are plots of $HR$ against absorption-corrected luminosity.  For
sources too weak for spectral analysis, the Galactic column density is
used in Fig.~\ref{fig4}a and the Galactic plus average intrinsic
column is used in Fig.~\ref{fig4}b.  It is notable that essentially
all sources with $2 \times 10^{38}$ erg s$^{-1} < L(0.4-8 \, \rm keV)
< 5 \times 10^{38}$ erg s$^{-1}$ have hard spectra, with $-0.5 <
\Gamma < 2.5$ (Fig.~\ref{fig4}b).

Under the assumption of absorption by the Galactic column density
only, 5 and 15 sources have luminosities (after correcting for
absorption) exceeding $10^{39}$ and $5 \times 10^{38}$ erg s$^{-1}$ in
the $0.4$--$8$~keV band, respectively.  Alternatively, if an intrinsic
column density of $N_{\rm H} = 1.28 \times 10^{21}$~cm$^{-2}$ is
assumed, then 5 and 18 sources have luminosities (after correcting for
absorption) exceeding $10^{39}$ and $5 \times 10^{38}$ erg s$^{-1}$ in
the $0.4$--$8$~keV band, respectively.

\subsection{The Distribution of Source Luminosities}
\label{sec-xlf}

The distribution of luminosities of the sources that we detect on chip
S3 in the NGC 1068 observation is shown in Fig.~\ref{fig5}.  We have
excluded from this figure the three sources that were undetected in
the $0.4$--$5$~keV band image (marked by open circles with upward and
rightward pointing arrows in Fig.~\ref{fig4}).  Also shown in
Fig.~\ref{fig5} is the distribution of luminosities (assuming their
distances to be that of NGC 1068) for the sources that we detect on
chip S3 in the 3C~47 observation (see \S~\ref{sec-detect}); we
consider these sources as representative of the unrelated foreground
and background objects in the NGC 1068 field.  Most of the sources in
the 3C~47 field have insufficient counts for a spectral analysis, and
so we have assumed a power-law continuum (of photon index $\Gamma =
2.0$, typical of an AGN) modified by the Galactic column density
towards 3C~47 of $N_{\rm H} (\rm Gal) = 5.87 \times 10^{20}$~cm$^{-2}$
\citep{dl90}.  This spectrum was then used to estimate fluxes in the
$0.4$--$8$~keV band from the number of counts observed in the
$0.4$--$5$~keV band.  We detect two sources in the soft
($0.4$--$1.5$~keV) band image of 3C~47 that were undetected in both
the full ($0.4$--$5$~keV) and hard ($1.5$--$5$~keV) band images, and
so we have used the number of counts observed in the soft band to
estimate the flux in the $0.4$--$8$~keV band.  For luminosities
greater than $\sim 10^{38}$~erg~s$^{-1}$, the surface density of
sources in the 3C~47 field is much lower than that towards NGC 1068,
so we can ignore the contribution of unrelated foreground and
background objects to the luminous source population in NGC 1068.
There is, however, one source in the 3C~47 field with a luminosity of
$>10^{39}$~erg~s$^{-1}$ in the $0.4$--$8$~keV band at the distance of
NGC 1068.  This source has a flux of $1.3 \times 10^{-13}$ erg
cm$^{-2}$ s$^{-1}$ in the $2$--$10$~keV band (assuming a power-law of
photon index $\Gamma = 2.0$ and the Galactic absorption towards 3C~47
of $N_{\rm H} (\rm Gal) = 5.87 \times 10^{20}$~cm$^{-2}$).  At this
flux level, we would expect $\sim 0.2$ sources in the 3C~47 field,
from the results of the \emph{ASCA} Medium-Sensitivity Survey
\citep{ued99}, and so there is a $\sim 20$\% chance of detecting a
foreground or background source this bright in the NGC 1068 field.

The observed distribution of source luminosities in NGC 1068 is
subject to uncertainty due to the Poissonian nature of observed source
counts, source confusion, incompleteness, and contamination by sources
that are spurious.  While Monte Carlo simulations can be used to
estimate the effect of these biases (e.g., Zezas \& Fabbiano 2002; Kim
\& Fabbiano 2002), this approach is unlikely to benefit us here, since
there are 54 sources in Table~\ref{tbl-1} with a signal-to-noise ratio
$\leq 5$ as a result of diffuse emission at small galactocentric
distances in NGC 1068.  We do not consider it worthwhile computing the
cumulative XLF, since the error in the slope is likely to be much
larger than the distribution of slopes presented in \citet{kil02}.

\subsection{Source Variability}
\label{sec-varability}

The brightest source, J024238.9$-$000055, is possibly variable, with a
$0.36$\% probability that the arrival times of the source events are
drawn from the same parent population as those of the background
events (and that the $K$--$S$ statistic exceeds the observed value by
chance).  The light curve for this source is shown in Fig.~\ref{fig6}.
Although we cannot reject at $\geq 99$\% confidence, using a
$\chi^{2}$ test, the null hypothesis that the X-ray flux is constant,
there is a suggestion of a rapid decrease in the source flux some
$30$~ks into the observation, where the count rate decreases by a
factor of two in $\sim 2$~ks.

In order to search for long term variability we compared our data with
the \emph{Chandra} grating observations.  Due to the much poorer
signal-to-noise ratio in the grating data, we concentrated on the two
brightest sources.  In the High Energy Transmission Grating
Spectrometer (HETGS) observation performed on December 4--5, 2000
(obsid 332) for 46~ks, we detected $49.9 \pm 8.5$ and $43.0 \pm 7.7$
counts (after background subtraction) in the $0.4$--$5$~keV band from
J024238.9$-$000055 and J024237.9$-$000118, respectively.  We have also
modeled the spectra of both sources with the corresponding power-law
model given in Table~\ref{tbl-2}, keeping all the parameters fixed at
their best-fit values except for the power-law normalization which was
allowed to vary.  The observed fluxes in the $0.4$--$8$~keV band are
$1.0 \times 10^{-13}$ and $3.9 \times 10^{-14}$~erg~cm$^{-2}$~s$^{-1}$
for J024238.9$-$000055 and J024237.9$-$000118, respectively.  These
results suggest that the observed flux from J024238.9$-$000055 has
declined significantly (by a factor of $\sim 5$) between the ACIS-S
and HETGS observations, while the flux observed from
J024237.9$-$000118 has remained essentially unchanged.

The Low Energy Transmission Grating Spectrometer (LETGS) observation
was performed $\sim 1$~day later on December 5-6, 2000 (obsid 329) for
77~ks.  Source J024237.9$-$000118 was outside the field of view of the
zeroth order image, so we have only analyzed source
J024238.9$-$000055.  We detected $49.6 \pm 8.7$ counts (after
background subtraction) in the $0.4$--$5$~keV band.  A similar
spectral modeling approach was adopted for the LETGS spectrum as that
used for the spectrum from the HETGS observation.  We find the
observed flux in the $0.4$--$8$~keV band to be $1.1 \times
10^{-13}$~erg~cm$^{-2}$~s$^{-1}$, similar to that observed in the
HETGS observation.

\section{DISCUSSION}
\label{sec-discussion}

\subsection{The Intermediate Luminosity X-ray Objects (IXOs)} 
\label{sec-ixos}

We detect a total of 84 compact, non-nuclear X-ray sources on the S3
chip in our observation of NGC 1068.  After correcting for only
Galactic absorption, 5 of these sources have $0.4$--$8$~keV
luminosities greater than $10^{39}$~erg~s$^{-1}$, assuming that their
emission is isotropic and that they are located within NGC 1068.  All
of these sources project inside the 25.0 B-magnitude (arc sec)$^{-2}$
isophote of the galactic disk.  In Fig.~\ref{fig7} we show an optical
continuum image of NGC 1068 at $6100$~{\AA} \citep{pd93} with the
positions of 4 of these sources superposed [the other source is
located outside the $2^{\prime}\!.0 \times 2^{\prime}\!.0$ ($8.4
\times 8.4$~kpc) field of view].  Also shown in this image are the
positions of the sources which have luminosities (corrected for
absorption) between $5 \times 10^{38}$~erg~s$^{-1}$ and
$10^{39}$~erg~s$^{-1}$ in the $0.4$--$8$~keV band.  There are 3
sources in this luminosity range outside the field of Fig.~\ref{fig7}.
Two of these project outside the 25.0 B-magnitude (arc sec)$^{-2}$
isophote of the galactic disk.\footnote{There are 11 sources marked in
Fig.~\ref{fig7} with luminosities between $5 \times
10^{38}$~erg~s$^{-1}$ and $10^{39}$~erg~s$^{-1}$ (assuming an
association with NGC 1068).  We detect 10 sources in this luminosity
range on chip S3 assuming only Galactic absorption, and 13 sources if
both Galactic and internal absorption is assumed.}  Their hardness
ratios are $HR = 2.0 \pm 0.9$ and $HR = 0.33 \pm 0.06$, respectively.
For a source with a power-law spectrum of photon index $\Gamma = 2.0$,
typical of an AGN, $HR = 0.34$ if we assume absorption from the
Galactic column density only, and $HR = 0.51$ if a column density of
$N_{\rm H} = 1.28 \times 10^{21}$~cm$^{-2}$ is assumed.  Thus, we
conclude that the spectrum of one of these sources is consistent with
that of a typical AGN, while the spectrum of the other source is
consistent with that of a typical AGN only if it is highly absorbed.

If the 5 sources with $0.4$--$8$~keV luminosities greater than
$10^{39}$~erg~s$^{-1}$ are X-ray binaries accreting with luminosities
that are sub-Eddington (i.e., $L_{\rm x} \simless 1.5 \times 10^{38}
\, (M/M_{\odot})$~erg~s$^{-1}$, where $M$ and $M_{\odot}$ are the
masses of the X-ray source and the Sun, respectively), then the
implied source masses are $\simgreat 7$--$100 \, M_{\odot}$.  Such
objects are almost certainly black holes, since the upper limit on the
mass of a non-rotating neutron star is $\sim 3 \, M_{\odot}$ (for a
review see e.g., van Paradijs \& McClintock 1995).  The maximum mass
of a black hole formed by the collapse of a single, non-zero
metallicity star within the mass range $\sim 25$--$40 \, M_{\odot}$ is
commonly believed to be $\sim 15 \, M_{\odot}$ due to the effects of
mass loss through stellar winds.  The black hole mass can be even
lower ($\simless 10 \, M_{\odot}$) for black hole progenitors in close
separation binary systems since, prior to collapse, the progenitor's
helium core may undergo mass loss in a Wolf-Rayet phase if the
hydrogen layers are removed by the companion star in a common envelope
phase \citep{fk01}.  However, the limit on the black hole mass may be
higher, since stars more massive than $\sim 40 \, M_{\odot}$ may not
produce a supernova explosion and all of the stellar material falls
back onto the black hole \citep{fry99}.  Moreover, the mass loss rates
for low metallicity stars may be small, thus allowing the black hole
progenitors to retain most of their original stellar mass (Vink, de
Koter, \& Lamers 2001).  Stars with masses within the range $\sim
100$--$250 \, M_{\odot}$ may undergo explosive nuclear burning at the
end of their evolution, however, and no remnant is left (Woosley \&
Weaver 1982; Glatzel, El Eid, \& Fricke 1985; Woosley 1986).
Alternatively, black holes with masses $\simgreat 100 \, M_{\odot}$
may originate from an earlier population of stars with effectively
zero metallicity (i.e., Population III stars) and masses $\simgreat
250 \, M_{\odot}$ (Woosley \& Weaver 1982; Bond, Arnett, \& Carr 1984;
Fryer, Woosley, \& Heger 2001).  These stars may not experience
substantial mass loss during their lifetimes, and a massive black hole
can form inside the star before explosive burning reverses its
collapse \citep{bon84}.

\subsection{Accretion Disk Models}
\label{sec-xrbs}

While single component MCD and power-law continuum models have been
used in the analysis of other IXOs (e.g., Makishima et al. 2000;
Strickland et al. 2001), such an analysis may be too simple, leading
to incorrect physical insights.  The reader (and the authors) may
therefore have misgivings about the following discussion, which is
based on the results of the spectral analysis presented in
\S~\ref{sec-spectral}.

The disk bolometric luminosity, $L_{\rm bol}$, the inner disk radius,
$R_{\rm in}$, and the primary mass, $M_{\rm x}$, are given in
Table~\ref{tbl-3} for each of the spectrally modeled sources, assuming
the results of the MCD model and that the disk is seen at $\theta =
60^{\circ}$ [$L_{\rm bol} \propto (\cos\theta)^{-1}$ and $R_{\rm in}
\propto M_{\rm x} \propto (\cos\theta)^{-1/2}$, where $\theta$ is the
inclination of the disk (face on corresponds to $\theta = 0^{\circ}$);
see e.g., Makishima et al. (2000)].  The disk bolometric luminosities
and color temperatures for each of the spectrally modeled sources are
shown in Fig.~\ref{fig8}.  The lines of constant mass and Eddington
ratio were added using Equations (9) and (11) of \citet{mak00},
assuming $\xi = 0.412$, $\kappa = 1.7$, and $\alpha = 1$.  As can be
seen in this figure, the majority of the sources have color
temperatures similar to those observed in Galactic black hole
candidates (i.e., $kT_{\rm in} \simless 1.5$~keV; Tanaka 2000),
primary masses of a few to a few tens of $M_{\odot}$, and luminosities
which do not exceed the Eddington limit (i.e., they are consistent
with being sources similar to the black hole candidates seen in our
own Galaxy).  However, Fig.~\ref{fig8} shows three sources that differ
from these trends, as discussed below.

The disk luminosities and color temperatures of the spectrally modeled
sources are compared with the predictions of the slim disk model in
Fig.~\ref{fig9} (cf. Fig.~1 in Watarai et al. 2001).  Most of the
sources have mass accretion rates [$\dot{m} = \dot{M} / (L_{\rm Edd} /
c^{2})$] of $\dot{m} \simless 20$, and values of $M_{\rm x}$ similar
to those inferred from the MCD model.

\subsubsection{J024237.9$-$000118 and J024240.4$-$000053}

The MCD model gives color temperatures of $kT_{\rm in} \sim 0.3$~keV
for two of the spectrally modeled sources (J024237.9-000118 and
J024240.4-000053) which, together with the observed bolometric
luminosity, yields an inner disk radius of $R_{\rm in} \simgreat
10^{3}$~km and an estimated primary mass of $M_{\rm x} \simgreat 100
\, M_{\odot}$ for each source (Table~\ref{tbl-3}; assuming the disk to
be inclined at $60^{\circ}$).  These values are much larger than those
derived from the spectral modeling of other IXOs (e.g., Makishima et
al. 2000; but see Colbert \& Mushotzky 1999 for a possible exception)
and Galactic black hole candidates (e.g., Ebisawa et al. 1994; Dotani
et al. 1997; Kubota et al. 1998; \.{Z}ycki, Done, \& Smith 1999).  The
disk bolometric luminosities are of order $0.1 \, L_{\rm Edd}$ for
both sources (Fig.~\ref{fig8}), which is lower than that derived for
other IXOs (see our Fig.~\ref{fig8} and Fig.~3 in Makishima et
al. 2000), but consistent with estimated values of $L_{\rm bol}
\simless 0.1 \, L_{\rm Edd}$ for Galactic black hole candidates in the
low state (e.g., Nowak 1995).

\subsubsection{J024238.9$-$000055}

The measured color temperature and bolometric luminosity of the most
luminous source in NGC 1068, J024238.9$-$000055, gives an estimated
primary mass of $M_{\rm x} \sim 2 \, M_{\odot}$ (assuming the disk to
be inclined at $60^{\circ}$), which is much lower than that derived
assuming the source to be radiating at the Eddington limit.  Such low
values of $M_{\rm x}$ have been found in several other IXOs and the
Galactic superluminal source GRS~1915$+$105 (e.g., Makishima et
al. 2000).  \citet{miz99} and \citet{mak00} have argued that some IXOs
may contain Kerr (i.e., spinning) black holes, so allowing a higher
value of $M_{\rm x}$ (a factor of $6$ higher for a given value of
$R_{\rm in}$ and a maximally rotating black hole).  This would not
work in the case of J024238.9$-$000055, since the mass derived
assuming the source is radiating at the Eddington luminosity exceeds
the value given in Table~\ref{tbl-3} and Fig.~\ref{fig8} by a factor
$\simeq 100$.  Thus, we conclude that the MCD interpretation requires
the source to be radiating above the Eddington limit, where the
assumption of accretion via a geometrically thin disk is likely to be
invalid.  The slim disk model (Fig.~\ref{fig9}) also requires
J024238.9$-$000055 to be undergoing supercritical accretion, with
$\dot{m} \simgreat 10^{4}$, which is an implausibly high accretion
rate.

The alternative description of a power-law spectrum gives a photon
index of $\Gamma = 0.85^{+0.12}_{-0.11}$ for J024238.9$-$000055, which
is much smaller than that seen in other IXOs and Galactic black hole
candidates (e.g., Tanaka 2000).  At low mass accretion rates, $\dot{m}
\simless 10^{-4}$, models of advection dominated accretion flows
predict hard X-ray spectra, which are dominated by thermal
bremsstrahlung radiation from electrons at $\approx 10^{9} \, \rm K$
(e.g., Esin, McClintock, \& Narayan 1997).  However, there are two
problems with such a model: (1) at energies in the \emph{Chandra}
band, a $\sim 100 \, \rm keV$ thermal bremsstrahlung continuum has a
slope of $\Gamma \approx 1.3$--$1.4$, which is significantly larger
than that observed in J024238.9$-$000055; and (2) such a low mass
accretion rate would require a supermassive ($\simgreat 10^{6} \,
M_{\odot}$) black hole to produce the observed luminosity, which is
unlikely since dynamical friction would cause such a black hole
located in the galactic disk to sink towards the center of the galaxy
in less than a Hubble time (e.g., Tremaine, Ostriker, \& Spitzer
1975).  Strong absorption and hard power-law continua ($0.8 \simless
\Gamma \simless 1.5$) characterize the 2--10~keV spectra of highly
magnetized neutron star binary systems (e.g., White, Nagase, \& Parmar
1995), and the local accretion rate onto the neutron star can exceed
the Eddington limit by a factor of $\sim 100$, since the radiation
escapes laterally out of the accretion column \citep{kle96,mil96}.
However, the brightest objects of this class (e.g., the sources
A0535-668, LMC~X-4, and SMC~X-1 in the Magellanic Clouds) have
2--10~keV luminosities which peak below $\simeq 10^{39}$~erg~s$^{-1}$
(Liu, van Paradijs, \& van den Heuvel 2000), an order of magnitude
lower than that observed from J024238.9$-$000055.  Another possibility
is that the X-ray emission from J024238.9$-$000055 may be the result
of inverse Compton scattering of synchroton photons in a jet (e.g.,
Markoff, Falcke, \& Fender 2001), an idea that could be tested by
future coordinated radio and X-ray observations of this source.

\section{COMPARISON WITH OTHER GALAXIES}
\label{sec-galaxies}

It is of interest to compare the X-ray properties of the IXOs in
galaxies with different modes of active star formation, e.g.,
circumnuclear starbursts, starbursts induced by galaxy-galaxy
interactions, and star formation in ``grand design'' spirals.  
We therefore compare the luminosities of the X-ray point sources in
NGC 1068 with those in the Antennae \citep{zez02}, M82 \citep{mat01},
M51 (Terashima \& Wilson 2002a), the Circinus galaxy \citep{sw01}, NGC
3256 \citep{lir02}, and M101 \citep{pen01}.  These galaxies have all
been imaged on the ACIS-S3 chip, except for M82, which has been imaged
with both the High Resolution Camera and the ACIS-I array.  Unless
stated otherwise, the \emph{Chandra} count rates were converted to
unabsorbed $0.4$--$8$~keV fluxes using {\sc PIMMS} version 3.2d
\citep{muk93} and assuming a 5~keV thermal bremsstrahlung continuum
spectrum with absorption by the Galactic column density only.  The
values of the Galactic column density towards each of the galaxies in
our sample were taken from \citet{dl90}.  This model gives fluxes
similar to those used to estimate the luminosities in Fig.~\ref{fig4}a
for most of the spectrally unmodeled sources in NGC 1068
(cf. \S~\ref{sec-spectral}).  We have taken the average of the counts
from individual sources in cases where individual galaxies have been
observed on more than one occasion with \emph{Chandra}.  Following
\citet{zez02}, a column density of $N_{\rm H} = 2 \times
10^{21}$~cm$^{-2}$ was assumed for sources in the Antennae that are
detected in the medium ($2$--$4$~keV) band, but not detected in the
soft ($0.3$--$2$~keV) band (see their Table~2).  \citet{mt99} found
strong absorption ($N_{\rm H} \sim 10^{22}$~cm$^{-2}$) in the residual
\emph{ASCA} spectrum of M82, obtained by subtracting the spectrum of
the lowest intensity state from that of the highest intensity state,
which may indicate the presence of strong absorption in the spectrum
of the brightest source in M82 (X$41.4+60$ or CXOU
J095550.2$+$694047).  A bright compact source in M51 (CXOM51
J132950.7$+$471155) also shows strong absorption, corresponding to
$N_{\rm H} \sim 5 \times 10^{22}$~cm$^{-2}$ \citep{tw02b}.  Therefore,
intrinsic column densities of $N_{\rm H} = 10^{22}$ and $5 \times
10^{22}$~cm$^{-2}$ were assumed for these two sources, respectively.
An analysis of the remaining 83 compact sources which project onto the
disk of NGC 5194 (the main galaxy in M51) found objects that are
detected in the hard ($2$--$8$~keV) band, but not detected in the soft
($0.5$--$2$~keV) band (Terashima \& Wilson 2002a).  The hard band
counts for these sources were converted to unabsorbed fluxes in the
$0.4$--$8$~keV band, assuming a 5~keV thermal bremsstrahlung continuum
absorbed by a column density of $N_{\rm H} = 2 \times
10^{21}$~cm$^{-2}$ which, as noted above, is the model assumed for the
sources in the Antennae that are not detected below $2$~keV.
\citet{sw01} report absorption in excess of the Galactic column for
several of the bright compact sources in the Circinus galaxy, and so
the $0.1$--$10$~keV count rates were converted to fluxes in the
$0.4$--$8$~keV band assuming the results of their spectral modeling.
The column densities from Table~3 in \citet{lir02} were adopted for
the compact sources detected in NGC 3256 (we excluded the southern
nucleus from our analysis since a low-luminosity AGN is likely to
contribute to its hard X-ray emission --- see Lira et al. for detailed
discussion).

The resulting $0.4$--$8$~keV fluxes were used to calculate source
luminosities, $L_{\rm x}$, assuming distances of 19.6~Mpc to the
Antennae ($H_{\rm o} = 75$~km~s$^{-1}$~Mpc$^{-1}$; Sandage \& Bedke
1994), 3.9~Mpc to M82 \citep{sm99}, 8.4~Mpc to M51 \citep{fel97},
$4.2$~Mpc to Circinus \citep{fre77}, $37.6$~Mpc to NGC 3256 ($H_{\rm
o} = 75$~km~s$^{-1}$~Mpc$^{-1}$; L\'ipari et al. 2000), and 7.2~Mpc to
M101 \citep{ste98}.  The total number and luminosity of sources with
$L_{\rm x} > 2.1 \times 10^{38}$~erg~s$^{-1}$ (i.e., sources with
luminosities exceeding the Eddington limit for a $1.4 \, M_{\odot}$
neutron star), observed $42.5$--$122.5 \mu$m luminosity, star
formation rate, and number of luminous X-ray sources divided by the
star formation rate are given in Table~\ref{tbl-4} for each of the
galaxies in our sample.  The observed $42.5$--$122.5 \mu$m luminosity
is calculated from
\begin{eqnarray}
L_{\rm fir} = 4 \pi D^{2} \, 1.26 \times 10^{-11} (2.58 \, f_{60} +
f_{100}) \, {\rm erg} \, {\rm s}^{-1}
\end{eqnarray}
(\emph{Cataloged Galaxies and Quasars Observed in the IRAS Survey},
1985, Appendix B), where $f_{60}$ and $f_{100}$ are the \emph{IRAS}
flux densities in Jy at $60\mu$m and $100\mu$m, respectively, and $D$
is the distance to the source in cm.  We have used the flux densities
given in the \emph{IRAS} Faint Source Catalog \citep{mos90}, except
for M82, NGC 5194, and M101, for which the flux densities were taken
from \citet{ric88}, and Circinus, for which the flux density was taken
from the \emph{IRAS Point Source Catalog} 1988.  Thermal radiation
from dust heated by massive O and B type stars dominates the
far-infrared emission between $42.5$ and $122.5 \mu$m in starburst and
spiral galaxies \citep{dy90}, and so $L_{\rm fir}$ should provide an
excellent measure of the formation rate of massive stars in our sample
of galaxies.  The star formation rate (column 5 of Table~\ref{tbl-4})
is calculated using Equation (26) in Condon (1992), and refers to
stars more massive than $\sim 5 \, M_{\odot}$.  X-ray binaries with
luminosities greater than $2.1 \times 10^{38}$~erg~s$^{-1}$ are likely
to have evolved from stars with initial masses greater than $\sim 8 \,
M_{\odot}$, so the formation rate of these stars should be similar
(Grimm, Gilfanov, \& Sunyaev 2003).  The \emph{Chandra} observation of
NGC 3256 was relatively short and so a large number of sources may
have gone undetected in this galaxy.  Further, NGC 3256 is much more
distant than the other galaxies, so only the more luminous X-ray
sources in this galaxy will have been detected.  For this galaxy, we
give a lower limit to the number of sources with luminosities
exceeding $2.1 \times 10^{38}$ erg s$^{-1}$.  In fact, we note that
all of the sources in NGC 3256 have unabsorbed $0.4$--$8$~keV
luminosities greater than $10^{39}$~erg~s$^{-1}$, and so this galaxy
has a larger number of IXOs than the Antennae.  The galactic disk of
M101 at the 25.0 B-magnitude (arc sec)$^{-2}$ isophote level subtends
a much larger solid angle than does the S3 chip, and so the number of
observed sources with luminosities exceeding $2.1 \times 10^{38}$ erg
s$^{-1}$ represents a lower limit to the actual number.  We have
ignored the possibility that some of the luminous X-ray sources may be
low mass X-ray binaries; such sources would be unrelated to recent
star formation and their presence would tend to decrease $N$.  This
effect is significant in the galaxies M101, Circinus, and NGC 5194,
which have stellar masses (see e.g., Grimm et al. 2003 and references
therein; Allen et al. 1978; Freeman et al. 1977) comparable to that of
the Milky Way, and massive star formation rates a few times higher
than that of the Milky Way (e.g., Grimm et al. 2003).  We estimate
that the fraction of sources in low mass systems with luminosities
exceeding $2.1 \times 10^{38}$~erg~s$^{-1}$ is between $\sim 25$ and
$\sim 50$\% for these galaxies, assuming that the formation rate of
such systems scales with the stellar mass of each galaxy.

The right-hand column of Table~\ref{tbl-4} shows that, for all
galaxies, except M82 and possibly NGC 3256, the value of
$\left[N(L_{\rm x} > 2.1 \times 10^{38} \, {\rm erg} \, {\rm
s}^{-1})/SFR(M_{\odot} \, {\rm yr}^{-1}) \right] = 10$ to within a
factor of two.  There are considerable observational uncertainties in
the measurement of both $N(L_{\rm x} > 2.1 \times 10^{38} \, {\rm
erg})$ and the $SFR$.  The value of $N$ is sensitive to absorption
within the galaxy, and the low rate for M82 may be, in part, a result
of absorption in this highly inclined galaxy.  The low $N/SFR$ ratio
in NGC 3256 is obviously related to its much greater distance than the
other galaxies, with many X-ray sources having $L_{\rm x} > 2.1 \times
10^{38} \, {\rm erg} \, {\rm s}^{-1}$ being undetected.  At the
moment, it seems premature to make any statements about variations in
the ratio of the number of luminous X-ray sources to the star
formation rate.

\section{CONCLUSIONS}

The main results of our analysis of the compact X-ray source
population in the Seyfert 2 galaxy NGC 1068 are as follows: 1) We
detect a total of 84 compact sources on the S3 chip, of which 66 are
located within the $25.0$ B-magnitude (arc sec)$^{-2}$ isophote of the
galactic disk in NGC 1068.  2) After correcting for absorption, five
of these sources have $0.4$--$8$~keV luminosities greater than
$10^{39}$~erg~s$^{-1}$, assuming that their emission is isotropic and
that they are associated with NGC 1068, well in excess of the
Eddington limit for a $1.4 \, M_{\odot}$ neutron star, and so we refer
to these sources as IXOs.  If these five sources are X-ray binaries
accreting with luminosities that are both sub-Eddington and isotropic,
then the implied source masses are $\simgreat 7$--$100 \, M_{\odot}$.
3) The brightest compact source in NGC 1068 has a spectrum which is
much harder than that found in Galactic black hole candidates and
other IXOs, which may be the result of inverse Compton scattering of
synchrotron photons in a jet, or other physical processes, such as
complex absorption.  4) The brightest source also shows large
amplitude variability, with the flux possibly decreasing by a factor
of two in $\sim 2$~ks during our \emph{Chandra} observation, and by a
factor of five between our observation and the grating observations
taken just over nine months later.  5) The ratio of the number of
sources with luminosities greater than $2.1 \times
10^{38}$~erg~s$^{-1}$ in the $0.4$--$8$~keV band to the rate of
massive ($> 5 \, M_{\odot}$) star formation is the same, to within a
factor of two, for NGC 1068, the Antennae, NGC 5194 (the main galaxy
in M51), and the Circinus galaxy.  This suggests that the production
of X-ray binaries (and SNRs) per massive star is approximately the
same value for galaxies undergoing different types of current star
formation activity.

We thank Yuichi Terashima for providing advice on the data analysis
and results prior to publication, Sylvain Veilleux for useful
discussions, Cole Miller for a critical reading of the manuscript,
Ken-ya Watarai for making available his slim disk predictions, and
Yuxuan Yang for providing help with determining the limiting flux
level in our observation of NGC 1068.  We have made use of
photographic data obtained using the UK Schmidt Telescope, which was
operated by the Royal Observatory Edinburgh, with funding from the UK
Science and Engineering Research Council, until 1988 June, and
thereafter by the Anglo-Australian Observatory.  Original plate
material is copyright (c) the Royal Observatory Edinburgh and the
Anglo-Australian Observatory.  The plates were processed into the
present compressed digital form with their permission.  The Digitized
Sky Survey was produced at the Space Telescope Science Institute under
US Government grant NAG W$-$2166.  This research has made use of the
NASA/IPAC Extragalactic Database (NED) which is operated by the Jet
Propulsion Laboratory, California Institute of Technology, under
contract with the National Aeronautics and Space Administration, and
of NASA's Astrophysics Data System.  This research was supported by
NASA through grants NAG~81027 and NAG~81755 to the University of
Maryland.


\clearpage



\figcaption[f1.eps]{The \emph{Chandra} image of NGC 1068 on the scale
of the galactic disk in the $0.4$ to $5.0$~keV band.  The shading is
proportional to the logarithm of the intensity, and ranges from 0
(black) to 800 (white) counts pixel$^{-1}$.  The linear feature
running from P.A. $= 78^{\circ}$ through the nucleus to P.A. $=
258^{\circ}$ is a readout trail, and is not real. \label{fig1}}

\figcaption[f2.eps]{An optical image of NGC 1068 from the Digitized Sky
Survey blue plate showing the orientation of the S3 chip (white
square) with respect to the galactic disk of NGC 1068 at the $25.0$
B-magnitude (arc sec)$^{-2}$ isophote level (dashed line).  Also shown
in this image are the locations of the 84 sources (crosses) discussed
in \S~\ref{sec-detect} and listed in Table~\ref{tbl-1}. Coordinates
are for J2000.0. \label{fig2}}

\figcaption[f3.eps]{X-ray spectra of the two brightest compact X-ray
sources in NGC 1068: J024237.9$-$000118 (filled circles) and
J024238.9$-$000055 (crosses).  The data are shown together with the
folded model comprising the best-fit absorbed power-law
continuum. \label{fig3}}

\figcaption[f4.eps]{Luminosity dependence of the hardness ratio
$C(1.5$--$5.0~{\rm keV})/C(0.4$--$1.5~{\rm keV})$ for the X-ray
sources in NGC 1068.  Horizontal dashed lines correspond to photon
indices of $\Gamma = -1$, $0$, $1$, $2$, $3$, $4$, and $5$ when an
absorbed power-law model is assumed.  Filled squares correspond to
sources for which luminosities have been estimated from the results of
a spectral analysis (\S~\ref{sec-spectral}).  Filled circles depict
sources whose luminosities were estimated assuming the spectrum is an
absorbed power-law continuum (\S~\ref{sec-spectral}).  Open circles
and open squares represent sources detected in only the hard and soft
band, respectively.  Three sources were not detected in the whole band
(0.4--5.0~keV), were detected in the hard band, and were not detected
in the soft band (due to the presence of surrounding extended soft
emission).  These sources are represented by open circles with upward
and rightward pointing arrows.  (a) The Galactic column density is
assumed for sources too weak for spectral analysis.  (b) The Galactic
column density plus an intrinsic column density of $N_{\rm H} = 1.28
\times 10^{21}$~cm$^{-2}$ is assumed for sources too weak for spectral
analysis. \label{fig4}}

\figcaption[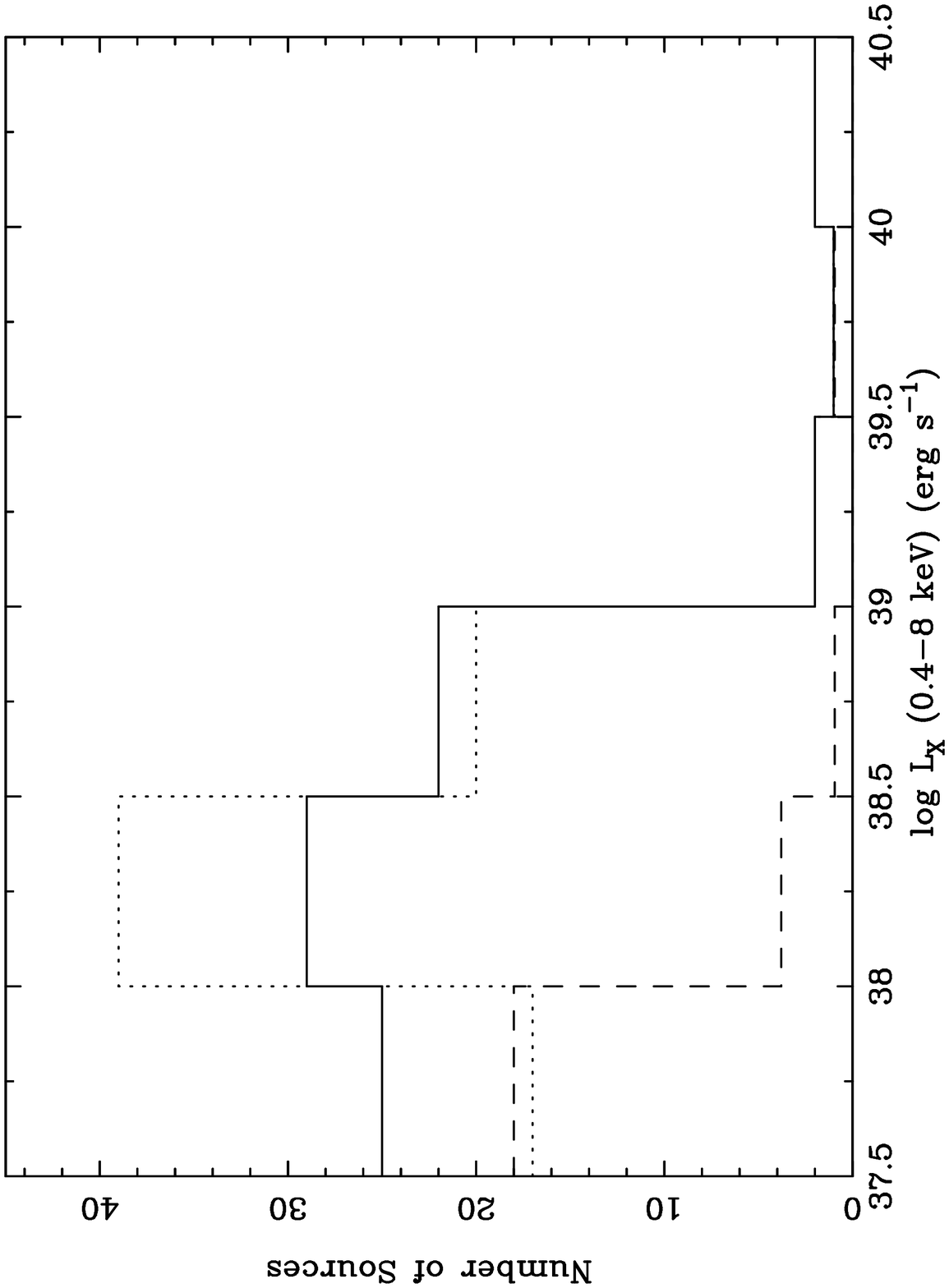]{The distribution of luminosities of the compact
X-ray sources, assuming they are associated with NGC 1068.
Luminosities were estimated either from the results of a spectral
analysis or assuming the spectrum is an absorbed power-law continuum
(\S~\ref{sec-spectral}).  In the latter case, the absorbing column was
taken to be either the Galactic column density towards NGC 1068 (solid
line) or the Galactic plus an intrinsic column density of $N_{\rm H} =
1.28 \times 10^{21}$~cm$^{-2}$ (dotted line).  The distribution of
luminosities of unrelated foreground and background sources (dashed
line) was estimated from the spectra of sources that we detect in the
field of 3C~47, assuming a power-law continuum model and Galactic
absorption. \label{fig5}}

\figcaption[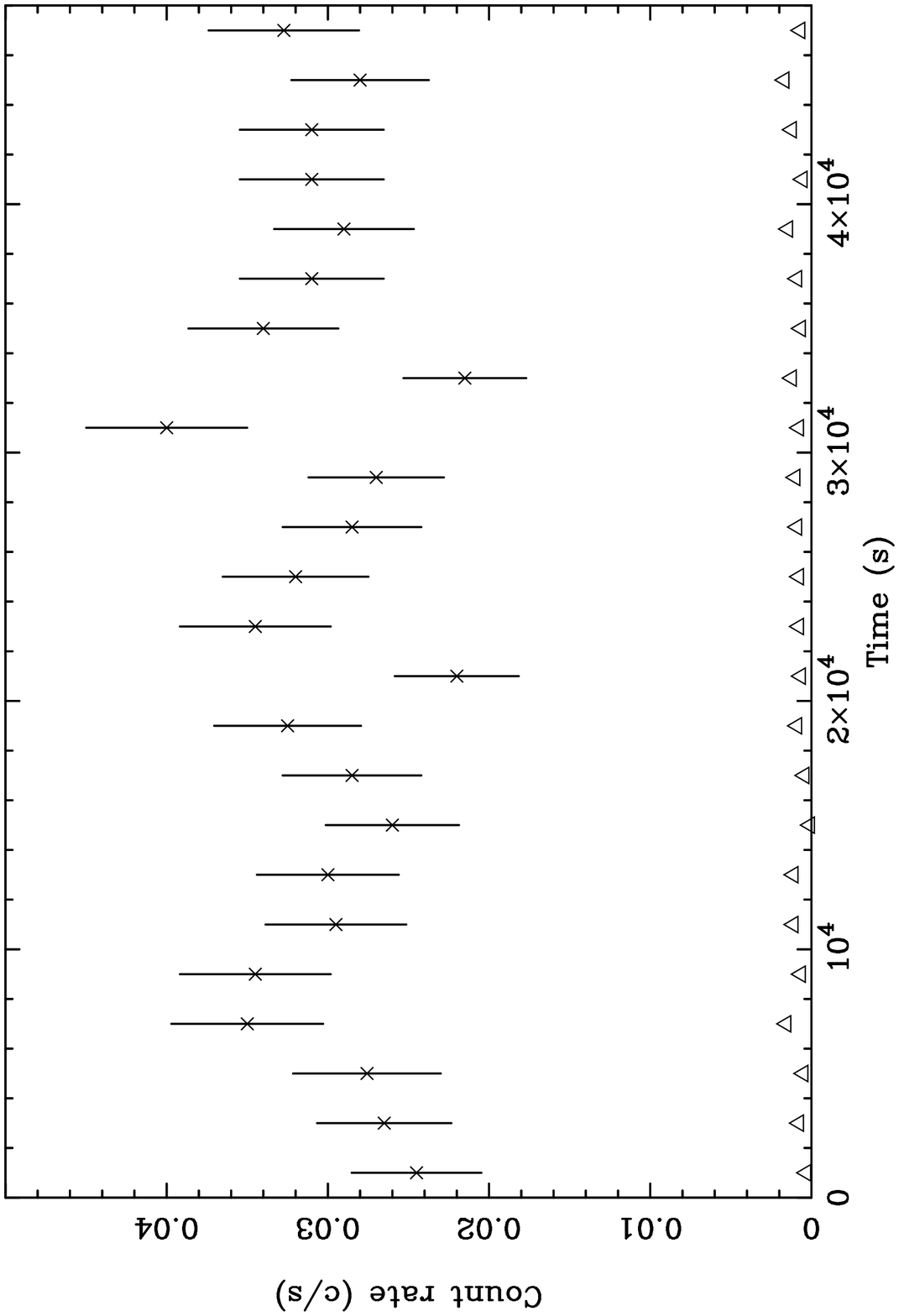]{The light curve for J024238.9$-$000055 binned so
that the duration of each data point is $2000$~s.  The crosses
represent the X-ray count rate within the source region, and include
both source and background photons.  Error bars correspond to $\pm
1\sigma$.  The open triangles correspond to the X-ray count rate
within the background region, but multiplied by the ratio of the
source to background areas.  The errors on the background rate are
comparable to the size of the triangles.
\label{fig6}}

\figcaption[f7.eps]{Optical continuum image of NGC 1068 at $6100$~{\AA}
(from Pogge \& De Robertis 1993).  The shading is proportional to the
logarithm of the intensity.  Crosses mark the locations of sources
with $0.4$--$8$~keV luminosities (after correcting for Galactic
absorption only; \S~\ref{sec-spectral}) greater than
$10^{39}$~erg~s$^{-1}$.  Asterisks identify sources with
$0.4$--$8$~keV luminosities (after correcting for Galactic absorption
only) in the range $5 \times 10^{38}$~erg~s$^{-1}$ to
$10^{39}$~erg~s$^{-1}$.  Open circles indicate 4 additional sources
with luminosities (after correcting for both the Galactic column
density and an intrinsic column density of $N_{\rm H} = 1.28 \times
10^{21}$~cm$^{-2}$; \S~\ref{sec-spectral}) greater than $5 \times
10^{38}$~erg~s$^{-1}$ but less than $10^{39}$~erg~s$^{-1}$.
\label{fig7}}

\figcaption[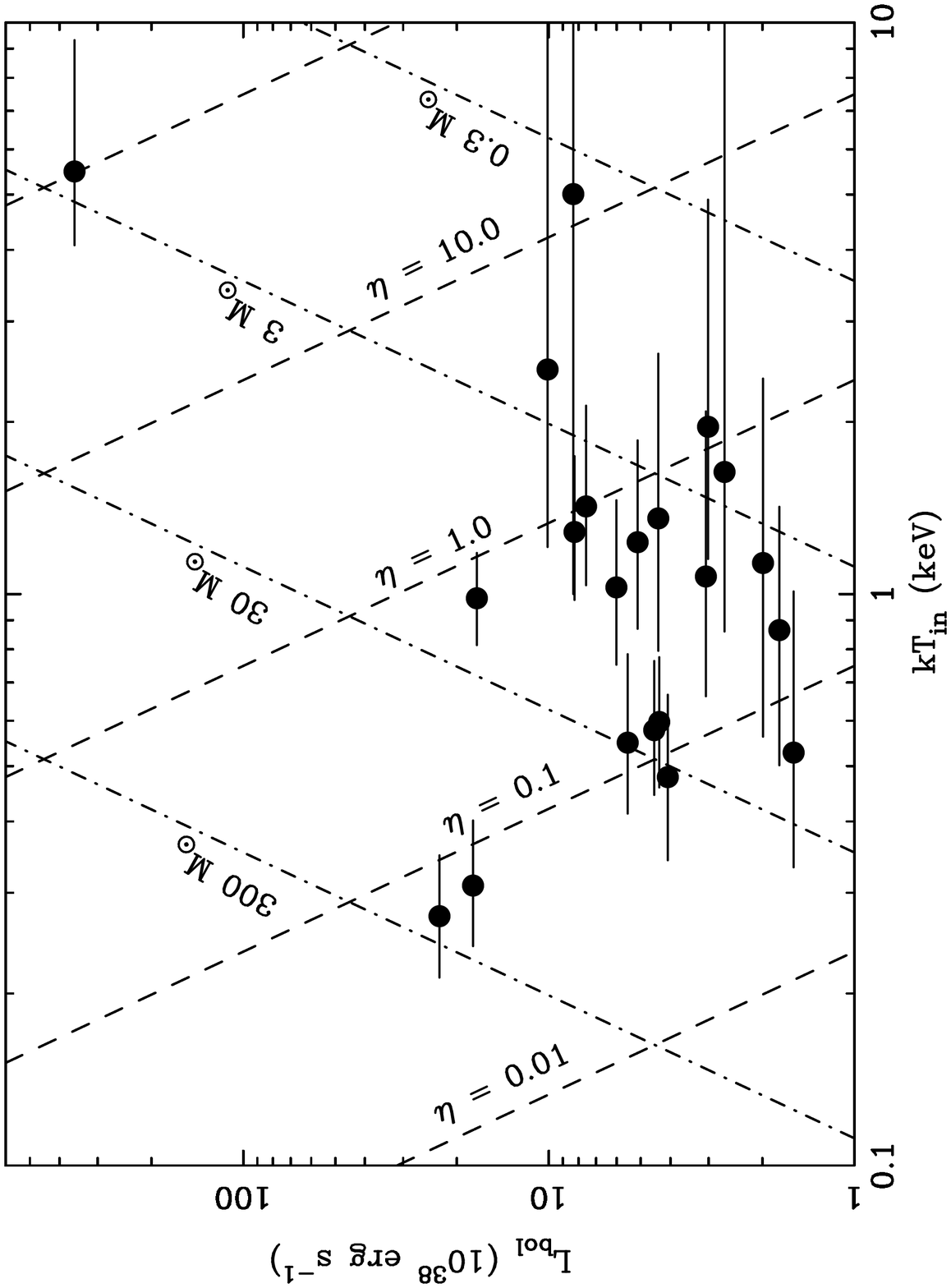]{The disk bolometric luminosity, $L_{\rm bol}$, is
plotted versus color temperature, $kT_{\rm in}$, for each of the
spectrally modeled sources in NGC 1068.  The straight lines represent
the predictions of the multi-color disk blackbody model of Makishima
et al. (2000).  Dash-dotted lines represent the loci of objects with
masses equal to $0.3$, $3$, $30$, and $300 \, M_{\odot}$.  Dashed
lines indicate the loci of objects with luminosities equal to $\eta =
0.01$, $0.1$, $1$, $10$, and $100$ times the Eddington luminosity,
according to this model. \label{fig8}}

\figcaption[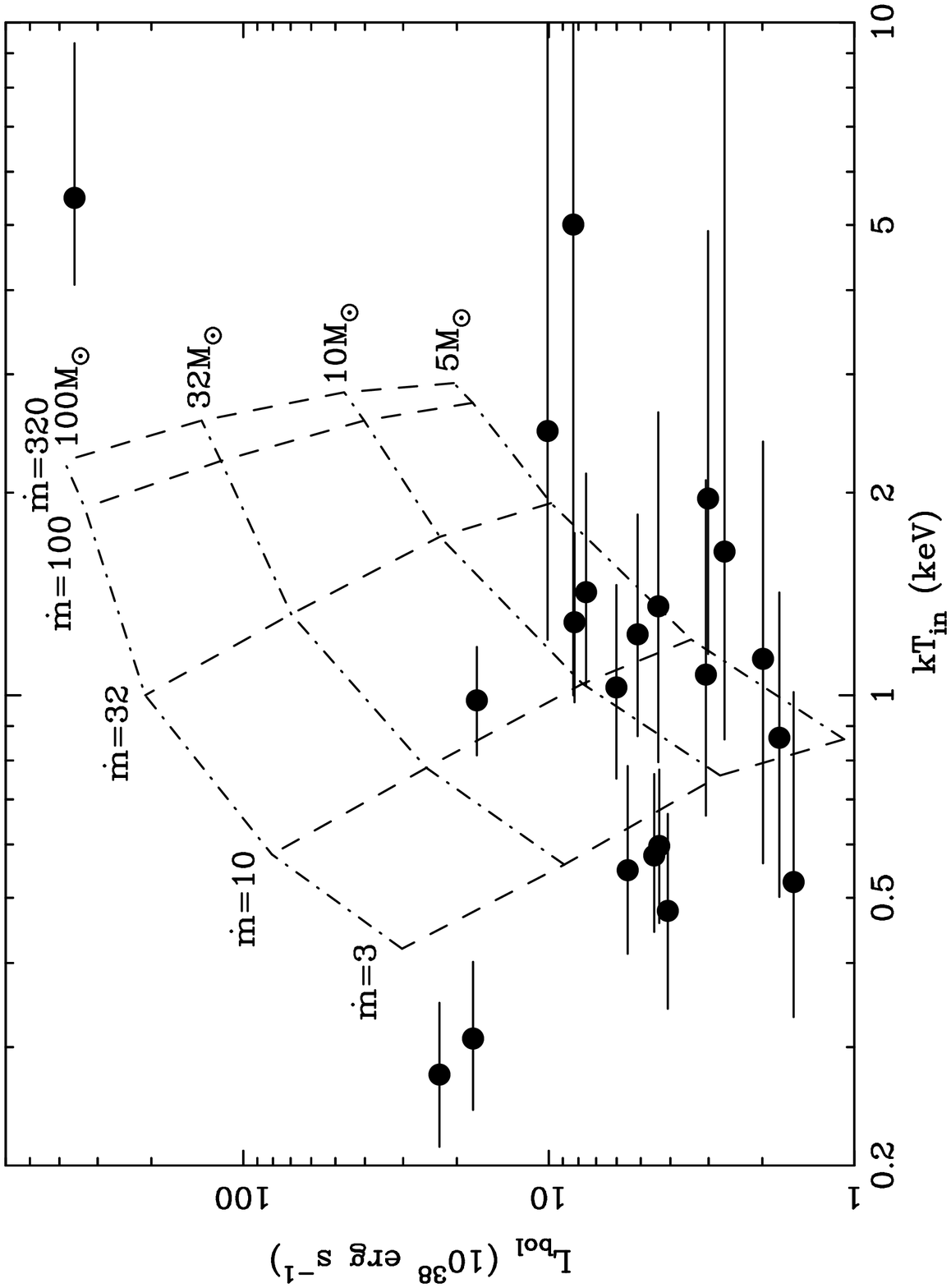]{The disk bolometric luminosity, $L_{\rm bol}$, is
plotted versus color temperature, $kT_{\rm in}$, for each of the
spectrally modeled sources in NGC 1068.  Dash-dotted lines represent
the loci of black holes with masses equal to $5$, $10$, $32$, and $100
\, M_{\odot}$.  Dashed lines indicate the loci of objects with
normalized mass accretion rates equal to $\dot{m} = 3$, $10$, $32$,
$100$, and $320$, according to the predictions of the slim disk model
of Watarai (2002). \label{fig9}}

\clearpage

\pagestyle{empty}

\begin{deluxetable}{cccccccc}
\footnotesize
\rotate 
\tablecaption{Sources in the NGC 1068 field. \label{tbl-1}}
\tablewidth{0pt}
\tablehead{
\colhead{Name} &
\colhead{} &
\colhead{} &
\colhead{Combined\tablenotemark{a}} &
\colhead{Soft\tablenotemark{b}} &
\colhead{Hard\tablenotemark{c}} &
\colhead{} &
\colhead{$\log L_{\rm X}$\tablenotemark{e}} \\
\colhead{CXOU} &
\colhead{R.A.(2000)} &
\colhead{Decl.(2000)} &
\colhead{Counts} &
\colhead{Counts} &
\colhead{Counts} &
\colhead{HR\tablenotemark{d}} &
\colhead{(erg s$^{-1}$)}}
\startdata 
J024224.5$+$000008 & 02 42 24.51 & $+$00 00 08.1 &    11.2 $\pm$ 4.6 &   7.6 $\pm$  4.0 &     3.6 $\pm$  3.2  & 0.5 $\pm$  0.5  & 37.6 \\     
J024227.5$-$000047 & 02 42 27.54 & $-$00 00 47.1 &    48.7 $\pm$ 8.2 &  42.1 $\pm$  7.6 &     6.6 $\pm$  3.8  & 0.2 $\pm$  0.1  & 38.1 \\
J024227.7$-$000202 & 02 42 27.67 & $-$00 02 02.0 &     6.8 $\pm$ 4.0 &   6.1 $\pm$  3.8 &             $<3.6$\tablenotemark{f} & $<0.6$ & 37.4 \\
J024228.2$+$000402 & 02 42 28.17 & $+$00 04 02.4 &    41.3 $\pm$ 8.0 &  13.7 $\pm$  5.3 &    27.7 $\pm$  6.7  & 2.0 $\pm$  0.9  & 38.4 \\  
J024228.5$-$000202 & 02 42 28.50 & $-$00 02 02.2 &    66.5 $\pm$ 9.3 &  38.5 $\pm$  7.3 &    28.0 $\pm$  6.4  & 0.7 $\pm$  0.2  & 38.4 \\
J024230.7$-$000030 & 02 42 30.65 & $-$00 00 29.8 &    26.8 $\pm$ 6.3 &   5.8 $\pm$  3.6 &    21.0 $\pm$  5.7  & 3.6 $\pm$  2.5  & 38.2 \\
J024232.8$+$000226 & 02 42 32.79 & $+$00 02 26.1 &    43.5 $\pm$ 7.8 &  27.7 $\pm$  6.5 &    15.8 $\pm$  5.1  & 0.6 $\pm$  0.2  & 38.2 \\
J024233.2$-$000105 & 02 42 33.23 & $-$00 01 04.9 &    61.3 $\pm$ 9.0 &  33.7 $\pm$  7.0 &    27.6 $\pm$  6.4  & 0.8 $\pm$  0.3 & 38.4 \\
J024233.9$-$000057 & 02 42 33.85 & $-$00 00 56.8 &    13.3 $\pm$ 5.1 &           $<5.4$\tablenotemark{f} & 11.7 $\pm$  4.6 & $>2.2$ & 37.9\\
J024234.1$-$000305 & 02 42 34.06 & $-$00 03 04.7 &    10.7 $\pm$ 4.5 &   5.0 $\pm$  3.4 &     5.7 $\pm$  3.7  & 1.1 $\pm$  1.1  & 37.6 \\
J024236.6$+$000013 & 02 42 36.64 & $+$00 00 13.2 &     9.6 $\pm$ 4.3 &   2.6 $\pm$  3.0 &     7.0 $\pm$  3.8  & 2.7 $\pm$  3.5  & 37.7 \\
J024236.7$-$000028 & 02 42 36.71 & $-$00 00 28.3 &    12.2 $\pm$ 5.0 &   8.5 $\pm$  4.4 &     3.7 $\pm$  3.3  & 0.4 $\pm$  0.4  & 37.5 \\
J024236.7$-$000109 & 02 42 36.73 & $-$00 01 09.2 &    40.0 $\pm$ 7.6 &  26.8 $\pm$  6.4 &    13.3 $\pm$  4.9  & 0.5  $\pm$  0.2 & 38.1  \\
J024237.0$+$000445 & 02 42 36.98 & $+$00 04 45.3 &    12.5 $\pm$ 5.3 &           $<9.4$\tablenotemark{f} & 7.8 $\pm$  4.3 & $>0.8$ & 37.7 \\
J024237.1$-$000019 & 02 42 37.10 & $-$00 00 18.7 &     9.1 $\pm$ 4.4 &           $<5.8$\tablenotemark{f} & 7.0 $\pm$  3.8 & $>1.2$ & 37.5 \\
J024237.7$-$000030 & 02 42 37.68 & $-$00 00 29.6 &    52.9 $\pm$ 8.6 &  23.4 $\pm$  6.3 &    29.6 $\pm$  6.6  & 1.3  $\pm$  0.4 & 38.3  \\
J024237.8$-$000143 & 02 42 37.84 & $-$00 01 43.4 &    31.1 $\pm$ 7.3 &  23.5 $\pm$  6.6 &     7.6 $\pm$  4.0  & 0.3  $\pm$  0.2 & 37.9  \\
J024237.9$-$000118 & 02 42 37.94 & $-$00 01 17.9 &   409 $\pm$ 22 & 348 $\pm$ 20 &    61.2 $\pm$  8.9  & 0.18  $\pm$  0.03 & 39.0 \\ 
J024238.9$-$000055 & 02 42 38.89 & $-$00 00 55.2 &  1334 $\pm$ 38 & 326 $\pm$ 20 &  1008 $\pm$ 33  & 3.1  $\pm$  0.2 & 39.9 \\
J024239.0$-$000050 & 02 42 38.96 & $-$00 00 50.0 &    36.0 $\pm$ 7.9 &  33.8 $\pm$  7.6 &             $<5.9$\tablenotemark{f} & $<0.2$ & 37.9 \\
J024239.0$-$000124 & 02 42 39.01 & $-$00 01 23.8 &    46.8 $\pm$ 9.4 &  41.5 $\pm$  9.1 &     5.4 $\pm$  3.7  & 0.13  $\pm$  0.09 & 38.0 \\
J024239.0$-$000057 & 02 42 39.05 & $-$00 00 56.6 &    44.0 $\pm$ 8.3 &  36.0 $\pm$  7.7 &     8.1 $\pm$  4.2  & 0.2  $\pm$  0.1 & 38.1 \\
J024239.3$-$000115 & 02 42 39.25 & $-$00 01 14.9 &    12.1 $\pm$ 7.1 &          $<11$\tablenotemark{f} &   8.5 $\pm$  4.4 & $>0.8$ & 37.7 \\
J024239.4$-$000035 & 02 42 39.39 & $-$00 00 35.4 &    73 $\pm$ 11 &  37.7 $\pm$  8.6 &    35.2 $\pm$  7.1  & 0.9  $\pm$  0.3 & 38.4 \\
J024239.4$-$000133 & 02 42 39.44 & $-$00 01 33.4 &     8.3 $\pm$ 4.5 &           $<6.9$\tablenotemark{f} &   5.7 $\pm$  3.6 & $>0.8$ & 37.5 \\
J024239.5$-$000104 & 02 42 39.48 & $-$00 01 03.6 &    47 $\pm$ 10 &  45 $\pm$ 10 &             $<6.2$\tablenotemark{f} & $<0.1$ & 38.1 \\
J024239.5$-$000028 & 02 42 39.54 & $-$00 00 28.2 &   177 $\pm$ 15 & 137 $\pm$ 13 &    40.5 $\pm$  7.6 & 0.30  $\pm$  0.06 & 38.7 \\
\\ \\ \\
J024239.7$-$000049 & 02 42 39.68 & $-$00 00 48.9 &            $<37$\tablenotemark{f} &          $<11$\tablenotemark{f} &    23.7 $\pm$  7.0 & $>2.1$ & 38.2 \\
J024239.7$-$000101 & 02 42 39.71 & $-$00 01 01.3 &   511 $\pm$ 26 & 341 $\pm$ 22 &   170 $\pm$ 14 & 0.50  $\pm$  0.05 & 39.2 \\
J024239.8$-$000039 & 02 42 39.80 & $-$00 00 39.4 &    55 $\pm$ 13 &  42 $\pm$ 12 &    13.3 $\pm$  5.6 & 0.3  $\pm$  0.2 & 38.2 \\
J024239.8$-$000107 & 02 42 39.82 & $-$00 01 07.1 &    23 $\pm$ 10 &  25 $\pm$ 10 &             $<3.2$\tablenotemark{f} & $<0.1$ & 37.7 \\
J024240.2$-$000054 & 02 42 40.16 & $-$00 00 53.6 &    38 $\pm$ 11 &          $<24$\tablenotemark{f} &    25.4 $\pm$  6.9 & $>1.1$ & 38.2 \\
J024240.4$-$000038 & 02 42 40.37 & $-$00 00 38.4 &    90 $\pm$ 15 &  80 $\pm$ 15 &    10.2 $\pm$  5.1  & 0.13  $\pm$  0.07 & 38.3 \\
J024240.4$-$000105 & 02 42 40.38 & $-$00 01 05.3 &            $<26$\tablenotemark{f} &          $<6.1$\tablenotemark{f} &    17.9 $\pm$  6.6 & $>2.9$ & 38.1 \\
J024240.4$-$000053 & 02 42 40.43 & $-$00 00 52.6 &   146 $\pm$ 19 & 127 $\pm$ 18 &    18.6 $\pm$  8.8  & 0.15  $\pm$  0.07 & 38.5 \\
J024240.5$-$000037 & 02 42 40.51 & $-$00 00 36.7 &   190 $\pm$ 19 & 157 $\pm$ 18 &    33.2 $\pm$  7.6  & 0.21  $\pm$  0.05 & 38.7 \\
J024240.5$+$000246 & 02 42 40.54 & $+$00 02 46.2 &     5.3 $\pm$ 3.6 &           $<2.3$\tablenotemark{f} &    5.5 $\pm$  3.6 & $>2.4$ & 37.5 \\
J024240.7$-$000132 & 02 42 40.68 & $-$00 01 31.8 &    23.4 $\pm$ 6.5 &   7.1 $\pm$  4.6 &    16.3 $\pm$  5.3  & 2.3  $\pm$  1.7 & 38.1 \\
J024240.7$-$000055 & 02 42 40.69 & $-$00 00 54.9 &   143 $\pm$ 18 &  93 $\pm$ 16 &    50.0 $\pm$  9.3  & 0.5  $\pm$  0.1 & 38.6 \\
J024240.7$-$000100 & 02 42 40.70 & $-$00 00 59.9 &    84 $\pm$ 12 &  25.8 $\pm$  9.0 &    58.2 $\pm$  9.1  & 2.2  $\pm$  0.9 & 38.7 \\
J024240.9$-$000144 & 02 42 40.86 & $-$00 01 43.6 &    34.3 $\pm$ 7.2 &   7.6 $\pm$  4.3 &    26.7 $\pm$  6.3  & 3.5  $\pm$  2.1 & 38.3 \\
J024240.9$-$000027 & 02 42 40.93 & $-$00 00 27.2 &   152 $\pm$ 18 &  91 $\pm$ 16 &    61.8 $\pm$  9.5  & 0.7  $\pm$  0.2 & 38.8 \\
J024241.0$-$000125 & 02 42 40.96 & $-$00 01 25.5 &    38.7 $\pm$ 8.1 &  24.0 $\pm$  7.0 &    14.7 $\pm$  5.0  & 0.6  $\pm$  0.3 & 38.2 \\
J024241.0$-$000051 & 02 42 40.97 & $-$00 00 51.0 &    64 $\pm$ 24 &  68 $\pm$ 23 &             $<7.2$\tablenotemark{f} & $<0.1$ & 38.2 \\
J024241.0$-$000038 & 02 42 40.98 & $-$00 00 38.0 &    80 $\pm$ 18 &  56 $\pm$ 17 &    24.4 $\pm$  7.1  & 0.4  $\pm$  0.2 & 38.3 \\
J024241.1$-$000039 & 02 42 41.14 & $-$00 00 39.4 &   136 $\pm$ 25 &  99 $\pm$ 24 &    36.3 $\pm$  8.9  & 0.4  $\pm$  0.1 & 38.6 \\
J024241.2$-$000040 & 02 42 41.19 & $-$00 00 39.8 &   126 $\pm$ 24 &  89 $\pm$ 23 &    37.0 $\pm$  8.7  & 0.4  $\pm$  0.1 & 38.5 \\
J024241.4$-$000057 & 02 42 41.37 & $-$00 00 57.1 &     9.0 $\pm$ 6.0 &           $<4.7$\tablenotemark{f} &    12.9 $\pm$  4.9 & $>2.7$ & 37.7 \\
J024241.4$-$000215 & 02 42 41.38 & $-$00 02 15.2 &    31.3 $\pm$ 6.8 &  12.6 $\pm$  4.7 &    18.6 $\pm$  5.5  & 1.5  $\pm$  0.7 & 38.1 \\
J024241.5$-$000253 & 02 42 41.47 & $-$00 02 53.2 &     8.7 $\pm$ 4.2 &   6.0 $\pm$  3.6 &             $<6.4$\tablenotemark{f} & $<1.1$ & 37.5 \\
J024241.6$-$000123 & 02 42 41.58 & $-$00 01 22.6 &    15.1 $\pm$ 5.8 &  14.7 $\pm$  5.6 &             $<3.4$\tablenotemark{f} & $<0.2$ & 37.6 \\
J024241.6$-$000110 & 02 42 41.61 & $-$00 01 09.6 &    12.6 $\pm$ 5.3 &   6.6 $\pm$  4.5 &     6.0 $\pm$  3.6  & 0.9  $\pm$  0.8 & 37.7 \\
J024241.7$+$000002 & 02 42 41.68 & $+$00 00 01.7 &    29.6 $\pm$ 7.8 &  24.3 $\pm$  7.2 &     5.3 $\pm$  4.0  & 0.2  $\pm$  0.2 & 37.9 \\
J024241.9$-$000028 & 02 42 41.88 & $-$00 00 28.5 &    39 $\pm$ 16 &          $<44$\tablenotemark{f} &    12.0 $\pm$  5.2 & $>0.3$ & 38.0 \\
J024241.9$-$000020 & 02 42 41.90 & $-$00 00 20.4 &    24 $\pm$ 12 &          $<29$\tablenotemark{f} &     9.0 $\pm$  4.4 & $>0.3$ & 37.8 \\
\\ \\
J024242.0$-$000136 & 02 42 41.98 & $-$00 01 35.7 &     5.2 $\pm$ 3.6 &           $<4.7$\tablenotemark{f} &     3.9 $\pm$  3.2 & $>0.8$ & 37.3 \\
J024242.2$-$000047 & 02 42 42.16 & $-$00 00 47.0 &            $<16$\tablenotemark{f} &          $<8.5$\tablenotemark{f} &     7.4 $\pm$  4.0 & $>0.9$ & 37.8 \\
J024242.5$-$000052 & 02 42 42.52 & $-$00 00 51.6 &    16.3 $\pm$ 6.2 &  15.0 $\pm$  6.0 &             $<4.7$\tablenotemark{f} & $<0.3$ & 37.7 \\
J024242.5$-$000048 & 02 42 42.54 & $-$00 00 48.1 &    24.6 $\pm$ 7.1 &          $<12$\tablenotemark{f} &    19.0 $\pm$  5.5 & $>1.6$ & 38.1 \\
J024242.6$+$000023 & 02 42 42.56 & $+$00 00 22.9 &    11.2 $\pm$ 5.0 &   5.5 $\pm$  4.1 &     5.8 $\pm$  3.6   & 1.1  $\pm$  1.0 & 37.6 \\
J024242.8$-$000245 & 02 42 42.78 & $-$00 02 44.8 &    57.4 $\pm$ 8.8 &  37.9 $\pm$  7.3 &    19.6 $\pm$  5.6   & 0.5  $\pm$  0.2 & 38.2 \\
J024242.8$-$000100 & 02 42 42.80 & $-$00 00 59.6 &     6.6 $\pm$ 4.4 &           $<5.3$\tablenotemark{f} &    5.6 $\pm$  3.6 & $>1.1$ & 37.4 \\
J024243.3$-$000141 & 02 42 43.34 & $-$00 01 40.7 &   117 $\pm$ 12 &  64.5 $\pm$  9.5 &    52.8 $\pm$  8.4   & 0.8  $\pm$  0.2 & 38.6 \\
J024244.0$-$000035 & 02 42 43.99 & $-$00 00 35.3 &    92 $\pm$ 11 &  42.4 $\pm$  8.5 &    49.4 $\pm$  8.2   & 1.2  $\pm$  0.3 & 38.5 \\
J024244.4$+$000435 & 02 42 44.40 & $+$00 04 35.1 &    23.9 $\pm$ 6.5 &   7.8 $\pm$  4.3 &    16.0 $\pm$  5.5   & 2.0  $\pm$  1.3 & 38.1 \\
J024244.6$+$000256 & 02 42 44.56 & $+$00 02 56.1 &    29.0 $\pm$ 6.5 &  16.0 $\pm$  5.1 &    13.0 $\pm$  4.8   & 0.8  $\pm$  0.4 & 38.0 \\
J024245.1$-$000010 & 02 42 45.12 & $-$00 00 10.0 &    63.6 $\pm$ 9.3 &  47.8 $\pm$  8.2 &    15.9 $\pm$  5.1   & 0.3  $\pm$  0.1 & 38.2 \\
J024245.4$+$000256 & 02 42 45.36 & $+$00 02 56.3 &    16.1 $\pm$ 5.5 &   9.8 $\pm$  4.6 &     6.2 $\pm$  3.9   & 0.6  $\pm$  0.5 & 37.8 \\
J024245.4$-$000219 & 02 42 45.43 & $-$00 02 19.0 &    54.0 $\pm$ 8.6 &  35.3 $\pm$  7.2 &    18.7 $\pm$  5.5   & 0.5  $\pm$  0.2 & 38.2 \\
J024245.5$-$000030 & 02 42 45.54 & $-$00 00 30.2 &    33.8 $\pm$ 7.5 &  12.1 $\pm$  5.5 &    21.7 $\pm$  5.8   & 1.8  $\pm$  0.9 & 38.3 \\
J024245.8$+$000006 & 02 42 45.79 & $+$00 00 06.0 &    19.4 $\pm$ 5.8 &   9.2 $\pm$  4.5 &    10.2 $\pm$  4.5   & 1.1  $\pm$  0.7 & 37.9 \\
J024247.4$+$000028 & 02 42 47.40 & $+$00 00 27.7 &   171 $\pm$ 14 &  26.7 $\pm$  6.6 &    38.8 $\pm$  7.3   & 1.5  $\pm$  0.5 & 38.8 \\
J024247.4$+$000351 & 02 42 47.41 & $+$00 03 50.7 &    36.3 $\pm$ 7.7 & 132 $\pm$ 13 &     9.7 $\pm$  4.8   & 0.07  $\pm$  0.04 & 37.9 \\
J024247.5$+$000501 & 02 42 47.54 & $+$00 05 00.5 &   226 $\pm$ 17 & 170 $\pm$ 14 &    55.9 $\pm$  9.0   & 0.33  $\pm$  0.06 & 38.8 \\
J024248.3$+$000329 & 02 42 48.32 & $+$00 03 29.4 &    22.9 $\pm$ 6.4 &  15.8 $\pm$  5.3 &     7.1 $\pm$  4.3   & 0.4  $\pm$  0.3 & 37.8 \\
J024248.6$+$000246 & 02 42 48.59 & $+$00 02 46.3 &    38.9 $\pm$ 7.9 &  26.1 $\pm$  6.8 &    12.8 $\pm$  5.0   & 0.5  $\pm$  0.2 & 38.0 \\
J024250.6$+$000546 & 02 42 50.58 & $+$00 05 45.8 &     9.1 $\pm$ 4.8 &           $<2.3$\tablenotemark{f} &    10.9 $\pm$  4.7 & $>4.7$ & 37.9 \\
J024250.9$+$000431 & 02 42 50.95 & $+$00 04 30.6 &    13.2 $\pm$ 5.2 &          $<8.4$\tablenotemark{f} &      9.1 $\pm$  4.4 & $>1.1$ & 37.7 \\
J024251.0$-$000032 & 02 42 50.99 & $-$00 00 31.6 &   207 $\pm$ 16 & 123 $\pm$ 12 &    84 $\pm$ 10   & 0.7  $\pm$  0.1 & 38.9 \\
J024252.0$+$000343 & 02 42 51.95 & $+$00 03 42.7 &    16.1 $\pm$ 5.8 &  11.4 $\pm$  4.9 &             $<9.4$\tablenotemark{f} & $<0.8$ & 37.8 \\
J024252.8$-$000055 & 02 42 52.76 & $-$00 00 55.5 &     9.5 $\pm$ 4.5 &   8.0 $\pm$  4.2 &             $<4.9$\tablenotemark{f} & $<0.6$ & 37.6 \\
J024253.1$-$000216 & 02 42 53.07 & $-$00 02 15.5 &    20.9 $\pm$ 5.8 &  17.2 $\pm$  5.4 &     3.7 $\pm$  3.2  & 0.2  $\pm$  0.2 & 37.8 \\
J024253.7$+$000309 & 02 42 53.70 & $+$00 03 08.7 &    25.7 $\pm$ 6.5 &  12.0 $\pm$  5.0 &    13.6 $\pm$  4.9  & 1.1  $\pm$  0.6 & 38.0 \\
\\ \\
J024255.9$-$000128 & 02 42 55.92 & $-$00 01 27.6 &    37.2 $\pm$ 7.3 &  29.0 $\pm$  6.5 &     8.2 $\pm$  4.3  & 0.3  $\pm$  0.2 & 38.0 \\
\enddata


\tablenotetext{a}{0.4--5.0~keV background subtracted counts.  The
errors here, and in columns 5 and 6, are $\sqrt{\left[ \sigma_{\rm
T}^{2} + \sigma_{\rm B}^{2} \times (A_{\rm T}/A_{\rm B})^{2} \right]}$
(e.g., Bevington \& Robinson 1992, p. 41), where $\sigma_{\rm T}$ and
$\sigma_{\rm B}$ are the respective $1\sigma$ errors on the total and
background counts, derived from equation (7) in Gehrels (1986), and
$A_{\rm T}$ and $A_{\rm B}$ are the respective areas of the source and
background regions.}
\tablenotetext{b}{0.4--1.5~keV background subtracted counts.}
\tablenotetext{c}{1.5--5.0~keV background subtracted counts.}
\tablenotetext{d}{Observed $C(1.5$--$5.0~{\rm keV})/C(0.4$--$1.5~{\rm
keV})$ hardness ratio.}
\tablenotetext{e}{0.4--5.0~keV luminosity corrected for Galactic
absorption only, assuming that the source is in NGC 1068 (distance $=
14.4$~Mpc) and the emission is isotropic.}
\tablenotetext{f}{Upper limits are calculated for a confidence level
of $90$\%, using either a) equation (9) in Kraft, Burrows, \& Nousek
(1991) for instances when the total number of counts in the detection
region is $\leq 20$, or b) equation (3) in Gehrels (1986), when the
total number of counts is $>20$.  The mean number of background counts
was then subtracted from these calculated upper limits.}

\end{deluxetable}

\clearpage

\begin{deluxetable}{cccccccc}
\footnotesize
\rotate
\tablecaption{Spectral Models of the Sources with $>50$ Counts (after Background Subtraction) and S/N $>7$.
\label{tbl-2}}
\tablewidth{0pt}
\tablehead{
\colhead{Source Name} &
\colhead{$N_{\rm H}$\tablenotemark{a}} &
\colhead{$\Gamma$\tablenotemark{b}} &
\colhead{$kT_{\rm in}$\tablenotemark{c}} &
\colhead{Norm\tablenotemark{d}} &
\colhead{$F_{\rm X}$\tablenotemark{e}} &
\colhead{$L_{\rm X}$\tablenotemark{f}} &
\colhead{$\chi^{2}$ (d.o.f.)} \\
\colhead{CXOU} &
\colhead{($10^{22}$ cm$^{-2}$)} &
\colhead{} &
\colhead{(keV)} &
\colhead{} &
\colhead{(erg cm$^{-2}$ s$^{-1}$)} &
\colhead{(erg s$^{-1}$)} &
\colhead{}}
\startdata
J024228.5-000202 & $<0.099$ & $1.36^{+0.75}_{-0.56}$ & ... &
$(1.36^{+0.47}_{-0.39}) \times 10^{-6}$ & $(1.06^{+0.75}_{-0.56})
\times 10^{-14}$ & $(2.7^{+1.8}_{-1.3}) \times 10^{38}$ & 9.5 (3) 
\\
J024228.5-000202 & $<0.094$ & ... & $1.6^{+\infty}_{-0.8}$ &
$(7^{+48}_{-7}) \times 10^{-5}$ & $(9.2^{+8.7}_{-4.6}) \times
10^{-15}$ & $(2.4^{+2.1}_{-1.1}) \times 10^{38}$ & 11.1 (3) \\
\\
J024233.2-000105 & $<0.16$ & $1.22^{+0.59}_{-0.36}$ & ... &
$(1.20^{+0.80}_{-0.35}) \times 10^{-6}$& $(1.10^{+0.31}_{-0.41})
\times 10^{-14}$ & $(2.8^{+0.8}_{-1.0}) \times 10^{38}$ & 2.5 (3) \\
J024233.2-000105 & $<0.10$ & ... & $2.0^{+2.9}_{-0.8}$ &
$(4^{+17}_{-4}) \times 10^{-5}$ & $(9.92^{+0.42}_{-0.40}) \times
10^{-15}$ & $(2.5^{+1.0}_{-1.0}) \times 10^{38}$ & 3.2 (3) \\
\\
J024237.7-000030 & $<0.36$ & $0.8^{+1.1}_{-0.5}$ & ... &
$(10^{+15}_{-3}) \times 10^{-7}$ & $(1.5^{+1.2}_{-0.9}) \times
10^{-14}$ & $(3.8^{+2.9}_{-1.9}) \times 10^{38}$ & 1.4 (2) \\
J024237.7-000030 & $<0.17$ & ... & $5.0^{+\infty}_{-4.0}$ &
$(3^{+336}_{-3}) \times 10^{-6}$ & $(1.4^{+0.3}_{-1.0})
\times 10^{-14}$ & $(3.6^{+0.7}_{-2.3}) \times 10^{38}$ & 1.5 (2) \\
\\
J024237.9-000118 & $0.44^{+0.20}_{-0.14}$ & $4.7^{+1.3}_{-0.9}$ &
... & $(4.4^{+3.8}_{-1.8}) \times 10^{-5}$ & $(2.78^{+0.56}_{-0.32})
\times 10^{-14}$ & $(8^{+57}_{-6}) \times 10^{39}$ & 30.1 (20) \\
J024237.9-000118 & $0.11^{+0.11}_{-0.09}$ & ... &
$0.309^{+0.092}_{-0.067}$ & $0.4^{+1.4}_{-0.3}$ &
$(2.59^{+0.27}_{-0.21}) \times 10^{-14}$ & $(1.1^{+1.0}_{-0.3})
\times 10^{39}$ & 33.7 (20) \\
\\
J024238.9-000055 & $0.51^{+0.13}_{-0.10}$ & $0.85^{+0.12}_{-0.11}$ &
... & $(4.01^{+0.81}_{-0.66}) \times 10^{-5}$ &
$(5.11^{+0.23}_{-0.27}) \times 10^{-13}$ & $(1.463^{+0.054}_{-0.058})
\times 10^{40}$ & 109.6 (90) \\
J024238.9-000055 & $0.472^{+0.073}_{-0.064}$ & ... &
$5.5^{+3.8}_{-1.4}$ & $(7^{+11}_{-6}) \times 10^{-5}$ &
$(4.98^{+0.29}_{-0.29}) \times 10^{-13}$ & $(1.408^{+0.047}_{-0.063})
\times 10^{40}$ & 106.8 (90) \\
\\
J024239.4-000035 & $0.21^{+0.43}_{-0.21}$ & $1.9^{+1.0}_{-0.8}$ &
... & $(3.3^{+5.0}_{-1.8}) \times 10^{-6}$ & $(1.24^{+0.63}_{-0.48})
\times 10^{-14}$ & $(4.2^{+7.3}_{-1.1}) \times 10^{38}$ & 3.1 (8)
\\
J024239.4-000035 & $0.06^{+0.30}_{-0.06}$ & ... & $1.1^{+1.0}_{-0.4}$
& $(4^{+27}_{-4}) \times 10^{-4}$ & $(1.00^{+0.62}_{-0.34})
\times 10^{-14}$ & $(2.8^{+1.3}_{-0.8}) \times 10^{38}$ & 3.0 (8) \\
\\
J024239.5-000028 & $0.056^{+0.099}_{-0.056}$ & $2.12^{+0.41}_{-0.57}$
& ... & $(5.7^{+4.5}_{-1.6}) \times 10^{-6}$ & $(2.11^{+0.95}_{-0.70})
\times 10^{-14}$ & $(6.4^{+5.9}_{-1.3}) \times 10^{38}$ & 8.0 (8)
\\
J024239.5-000028 & $<0.061$ & ... & $0.58^{+0.19}_{-0.13}$ &
$(8^{+16}_{-5}) \times 10^{-3}$ & $(1.38^{+0.33}_{-0.22}) \times
10^{-14}$ & $(3.76^{+0.76}_{-0.45}) \times 10^{38}$ & 10.1 (8) \\
\\
J024239.7-000101 & $0.127^{+0.076}_{-0.080}$ & $1.93^{+0.31}_{-0.23}$
& ... & $(1.96^{+0.69}_{-0.50}) \times 10^{-5}$ & $(7.6^{+1.5}_{-1.4})
\times 10^{-14}$ & $(2.42^{+0.41}_{-0.17}) \times 10^{39}$ & 31.2 (32)
\\
J024239.7-000101 & $<0.044$ & ... & $0.98^{+0.20}_{-0.17}$ &
$(3.4^{+3.6}_{-1.6}) \times 10^{-3}$ & $(6.0^{+1.1}_{-0.9}) \times
10^{-14}$ & $(1.56^{+0.26}_{-0.20}) \times 10^{39}$ & 33.0 (32) \\
\\
J024240.4-000053 & $0.79^{+0.30}_{-0.21}$ & $5.8^{+1.6}_{-1.1}$ &
... & $(6.8^{+8.8}_{-3.4}) \times 10^{-5}$ & $(1.53^{+0.16}_{-0.19})
\times 10^{-14}$ & $(2^{+32}_{-2}) \times 10^{40}$ & 22.9 (18)
\\
J024240.4-000053 & $0.35^{+0.20}_{-0.15}$ & ... &
$0.273^{+0.076}_{-0.060}$ & $0.77^{+4.21}_{-0.61}$ &
$(1.47^{+0.12}_{-0.15}) \times 10^{-14}$ & $(1.4^{+2.2}_{-0.6}) \times
10^{39}$ & 23.8 (18) \\
\\
\\ \\ \\
J024240.5-000037 & $0.09^{+0.12}_{-0.09}$ & $2.57^{+0.53}_{-0.88}$ &
... & $(6.0^{+8.2}_{-2.4}) \times 10^{-6}$ & $(1.59^{+0.87}_{-0.53})
\times 10^{-14}$ & $(6^{+28}_{-2}) \times 10^{38}$ & 7.9 (14) \\
J024240.5-000037 & $<0.10$ & ... & $0.48^{+0.19}_{-0.14}$ &
$(1.5^{+6.4}_{-1.1}) \times 10^{-2}$ & $(1.16^{+0.19}_{-0.18}) \times
10^{-14}$ & $(3.2^{+1.6}_{-0.2}) \times 10^{38}$ & 9.2 (14) \\
\\
J024240.7-000055 & $<0.14$ & $1.47^{+0.41}_{-0.28}$ & ... &
$(4.7^{+2.1}_{-1.0}) \times 10^{-6}$ & $(3.27^{+0.50}_{-0.70}) \times
10^{-14}$ & $(8.45^{+1.15}_{-1.35}) \times 10^{38}$ & 32.2 (20) \\
J024240.7-000055 & $<0.067$ & ... & $1.42^{+0.71}_{-0.39}$ &
$(3.4^{+6.9}_{-2.6}) \times 10^{-4}$ & $(2.66^{+0.70}_{-0.59}) \times
10^{-14}$ & $(6.8^{+1.7}_{-1.4}) \times 10^{38}$ & 35.1 (20) \\
\\
J024240.7-000100 & $0.90^{+1.38}_{-0.75}$ & $1.43^{+1.22}_{-0.91}$ &
... & $(5^{+19}_{-4}) \times 10^{-6}$ & $(2.66^{+0.87}_{-0.94})
\times 10^{-14}$ & $(10^{+30}_{-1}) \times 10^{38}$ & 13.9 (14) \\
J024240.7-000100 & $0.60^{+0.66}_{-0.41}$ & ... &
$3^{+\infty}_{-1}$ & $(5^{+70}_{-5}) \times 10^{-5}$ &
$(2.52^{+0.42}_{-0.92}) \times 10^{-14}$ & $(7.7^{+1.8}_{-2.0}) \times
10^{38}$ & 14.3 (14) \\
\\
J024240.9-000027 & $<0.092$ & $1.66^{+0.49}_{-0.28}$ & ... &
$(5.0^{+2.6}_{-1.0}) \times 10^{-6}$ & $(2.87^{+0.57}_{-0.76}) \times
10^{-14}$ & $(7.5 \pm 1.3) \times 10^{38}$ & 14.7 (16) \\
J024240.9-000027 & $<0.043$ & ... & $1.03^{+0.43}_{-0.28}$ &
$(1.0^{+2.0}_{-0.7}) \times 10^{-3}$ & $(2.10^{+0.62}_{-0.45})
\times 10^{-14}$ & $(5.5^{+1.5}_{-1.1}) \times 10^{38}$ & 19.7 (16) \\
\\
J024242.8-000245 & $<0.16$ & $1.59^{+0.76}_{-0.41}$ & ... &
$(1.5^{+1.1}_{-0.4}) \times 10^{-6}$ & $(9.2^{+3.2}_{-3.8})
\times 10^{-15}$ & $(2.39 \pm 0.76) \times 10^{38}$ & 1.2 (3) \\
J024242.8-000245 & $<0.064$ & ... & $1.1^{+1.3}_{-0.6}$ &
$(2^{+25}_{-2}) \times 10^{-4}$ & $(7.0^{+4.9}_{-3.2}) \times
10^{-15}$ & $(1.8^{+1.2}_{-0.7}) \times 10^{38}$ & 5.3 (3) \\
\\
J024243.3-000141 & $0.09^{+0.13}_{-0.09}$ & $1.49^{+0.81}_{-0.48}$ &
... & $(3.2^{+3.1}_{-1.2}) \times 10^{-6}$ & $(2.0^{+1.1}_{-0.8})
\times 10^{-14}$ & $(5.6^{+2.2}_{-1.4}) \times 10^{38}$ & 2.3 (4) \\
J024243.3-000141 & $<0.14$ & ... & $1.4^{+1.3}_{-0.6}$ &
$(2^{+15}_{-2}) \times 10^{-4}$ & $(1.6^{+1.2}_{-0.6}) \times
10^{-14}$ & $(4.0^{+2.9}_{-1.3}) \times 10^{38}$ & 2.6 (4) \\
\\
J024244.0-000035 & $0.42^{+0.33}_{-0.19}$ & $1.94^{+0.72}_{-0.41}$ &
... & $(5.5^{+5.9}_{-2.6}) \times 10^{-6}$ & $(1.74^{+0.30}_{-0.51})
\times 10^{-14}$ & $(6.8^{+6.4}_{-1.3}) \times 10^{38}$ & 3.0 (7) \\
J024244.0-000035 & $0.22^{+0.25}_{-0.15}$ & ... &
$1.23^{+0.62}_{-0.36}$ & $(4^{+13}_{-3}) \times 10^{-4}$ &
$(1.53^{+0.39}_{-0.37}) \times 10^{-14}$ & $4.7^{+0.9}_{-1.1})
\times 10^{38}$ & 2.4 (7) \\
\\
J024245.1-000010 & $<0.15$ & $2.0^{+1.4}_{-0.6}$ & ... &
$(1.8^{+1.7}_{-0.4}) \times 10^{-6}$ & $(8.0^{+5.7}_{-3.7}) \times
10^{-15}$ & $(2.1^{+3.8}_{-0.6} \times 10^{38}$ & 1.3 (3) \\
J024245.1-000010 & $<0.089$ & ... & $0.53^{+0.48}_{-0.20}$ &
$(4^{+26}_{-4}) \times 10^{-3}$ & $(4.7^{+4.5}_{-1.1}) \times
10^{-15}$ & $(1.3^{+1.1}_{-0.2}) \times 10^{38}$ & 4.2 (3) \\
\\
J024245.4-000219 & $0.26^{+0.34}_{-0.26}$ & $2.43^{+0.65}_{-0.93}$ &
... & $(2.8^{+4.5}_{-1.7}) \times 10^{-6}$ & $(6.3^{+1.1}_{-2.5})
\times 10^{-15}$ & $(3^{+12}_{-1}) \times 10^{38}$ & 0.51 (2) \\
J024245.4-000219 & $0.05^{+0.22}_{-0.05}$ & ... &
$0.86^{+0.56}_{-0.36}$ & $(6^{+59}_{-5}) \times 10^{-4}$ &
$(5.6^{+2.1}_{-2.0}) \times 10^{-15}$ & $(1.58^{+0.61}_{-0.49}) \times
10^{38}$ & 0.72 (2) \\
\\
\\ \\ \\
J024247.4+000028 & $0.028^{+0.094}_{-0.028}$ & $2.17^{+0.42}_{-0.36}$
& ... & $(5.0^{+3.0}_{-1.0}) \times 10^{-6}$ & $(1.89^{+0.41}_{-0.49})
\times 10^{-14}$ & $(5.5^{+3.2}_{-0.9}) \times 10^{38}$ & 2.7 (8)
\\
J024247.4+000028 & $<0.032$ & ... & $0.60^{+0.18}_{-0.14}$ &
$(6^{+12}_{-4}) \times 10^{-3}$ & $(1.35^{+0.25}_{-0.21}) \times
10^{-14}$ & $(3.65^{+0.57}_{-0.47}) \times 10^{38}$ & 9.7 (8) \\
\\
J024247.5+000501 & $<0.067$ & $2.04^{+0.39}_{-0.24}$ & ... &
$(6.1^{+1.9}_{-0.7}) \times 10^{-6}$ & $(2.64^{+0.46}_{-0.57})
\times 10^{-14}$ & $(7.1^{+1.3}_{-1.0}) \times 10^{38}$ & 5.1 (12)
\\
J024247.5+000501 & $<0.024$ & ... & $0.55^{+0.24}_{-0.14}$ &
$(1.1^{+2.4}_{-0.8}) \times 10^{-2}$ & $(1.66^{+0.48}_{-0.24})
\times 10^{-14}$ & $(4.5^{+1.1}_{-0.5}) \times 10^{38}$ & 21.6 (12)
\\
\\
J024251.0-000032 & $0.014^{+0.078}_{-0.014}$ & $1.40^{+0.28}_{-0.25}$
& ... & $(5.2^{+2.4}_{-0.9}) \times 10^{-6}$ &
$(3.9^{+0.9}_{-1.0}) \times 10^{-14}$ & $(1.01^{+0.21}_{-0.20})
\times 10^{39}$ & 4.8 (10) \\
J024251.0-000032 & $<0.034$ & ... & $1.28^{+0.46}_{-0.31}$ &
$(5.7^{+8.3}_{-3.6}) \times 10^{-4}$ & $(2.92^{+0.92}_{-0.72}) \times
10^{-14}$ & $(7.5^{+2.3}_{-1.8}) \times 10^{38}$ & 6.3 (10) \\
\enddata


\tablecomments{The errors given here are at the $90$\% confidence level for
one interesting parameter (i.e., $\Delta\chi^{2} = 2.706$).}

\tablenotetext{a}{Measured intrinsic column density.}
\tablenotetext{b}{Photon spectral index.}
\tablenotetext{c}{Inner disk temperature for multi-color disk
blackbody model.}
\tablenotetext{d}{Normalizations for the power-law and MCD models.
For the power-law model, the units are photons cm$^{-2}$ s$^{-1}$
keV$^{-1}$ at 1~keV.  For the MCD model, the normalization is equal to
$[(r_{\rm in}/{\rm km})/(D/10~\rm kpc)]^{2} \cos\theta$, where $r_{\rm
in}$ is the inner disk radius, $D$ is the distance to the source, and
$\theta$ is the inclination of the disk (face on corresponds to
$\theta = 0^{\circ}$).}
\tablenotetext{e}{Observed 0.4-8~keV flux.}
\tablenotetext{f}{0.4--8~keV luminosity corrected for both Galactic
and intrinsic absorption, assuming that the source is in NGC 1068
(assumed distance $= 14.4$~Mpc).  Luminosities are calculated assuming
the emission to be isotropic for the power-law model, and the disk to
be inclined at $60^{\circ}$ for the MCD model.}

\end{deluxetable}

\clearpage

\begin{deluxetable}{cccc}
\footnotesize
\tablecaption{MCD model parameters. \label{tbl-3}}
\tablewidth{0pt}
\tablehead{
\colhead{Source Name} &
\colhead{$L_{\rm bol}$\tablenotemark{a}} &
\colhead{$R_{\rm in}$\tablenotemark{b}} &
\colhead{$M_{\rm X}$\tablenotemark{c}} \\
\colhead{CXOU} &
\colhead{($\times 10^{38}$ erg s$^{-1}$)} &
\colhead{(km)} &
\colhead{($M_{\odot}$)}}
\startdata
J024228.5-000202 & $2.7$ & $17$  & $2.1$ \\
J024233.2-000105 & $3.0$ & $13$  & $1.5$ \\
J024237.7-000030 & $8.3$ & $3.2$ & $0.39$ \\
J024237.9-000118 & $18$  & $1200$ & $150$ \\
J024238.9-000055 & $360$ & $17$  & $2.2$ \\
J024239.4-000035 & $3.1$ & $42$  & $5.2$ \\
J024239.5-000028 & $4.5$ & $180$ & $22$ \\
J024239.7-000101 & $1.7$ & $120$ & $15$ \\
J024240.4-000053 & $23$  & $1800$ & $220$ \\
J024240.5-000037 & $4.1$ & $250$ & $30$ \\
J024240.7-000055 & $7.6$ & $38$  & $4.7$ \\
J024240.7-000100 & $10$  & $15$  & $1.8$ \\
J024240.9-000027 & $6.0$ & $65$  & $8.0$ \\
J024242.8-000245 & $2.0$ & $31$  & $3.8$ \\
J024243.3-000141 & $4.4$ & $32$  & $3.9$ \\
J024244.0-000035 & $5.1$ & $41$  & $5.1$ \\
J024245.1-000010 & $1.6$ & $130$ & $15$ \\
J024245.4-000219 & $1.8$ & $49$  & $6.1$ \\
J024247.4+000028 & $4.4$ & $160$ & $20$ \\
J024247.5+000501 & $5.5$ & $220$ & $27$ \\
J024251.0-000032 & $8.2$ & $48$  & $6.0$ \\
\enddata


\tablecomments{Quantities in the table scale as $L_{\rm bol} \propto
(\cos\theta)^{-1}$ and $M_{\rm X} \propto R_{\rm in} \propto
(\cos\theta)^{-1/2}$, where $\theta$ is the inclination of the disk.
We have assumed $\cos\theta = 0.5$, corresponding to the angle
averaged value, $\xi \, \kappa^{2} = 1.18$, and $\alpha = 1$
(\S~\ref{sec-xrbs}).}

\tablenotetext{a}{Disk bolometric luminosity calculated using equation
(3) in Makishima et al. (2000).}
\tablenotetext{b}{Inner disk radius, derived from equation (5) in
Makishima et al. (2000).}
\tablenotetext{c}{Mass of the compact object estimated from equation
(8) in Makishima et al. (2000).}

\end{deluxetable}

\clearpage

\begin{deluxetable}{cccccc}
\footnotesize
\rotate
\tablecaption{Luminosities of the Compact X-ray Sources in Nearby Galaxies. \label{tbl-4}}
\tablewidth{0pt}
\tablehead{
\colhead{Galaxy} &
\colhead{$N(L_{\rm x} > 2.1 \times 10^{38}$~erg~s$^{-1})$\tablenotemark{a}} &
\colhead{$\log$ ($L_{\rm x, total}$)\tablenotemark{b}} &
\colhead{$\log$ ($L_{\rm fir}$)\tablenotemark{c}} &
\colhead{$SFR$\tablenotemark{d}} &
\colhead{$\left[\frac{N(L_{\rm x} > 2.1 \times 10^{38} \, {\rm erg} \, 
{\rm s}^{-1})}{SFR}\right]$}
\\
\colhead{} &
\colhead{} &
\colhead{(erg s$^{-1}$)} &
\colhead{(erg s$^{-1}$)} &
\colhead{($M_{\odot} \, \rm yr^{-1}$)} &
\colhead{}}
\startdata
NGC 1068     &  $24$ &  $40.2$ & $44.3$ &  $5.0$ &  $5.0$ \\
the Antennae &  $28$ &  $40.6$ & $44.0$ &  $2.4$ &  $12$ \\
M82          &   $6$ &  $40.7$ & $44.0$ &  $2.2$ &  $2.8$ \\
NGC 5194     &  $11$ &  $40.1$ & $43.6$ &  $0.9$ &  $12$ \\
Circinus     &   $9$ &  $40.4$ & $43.4$ &  $0.6$ &  $15$ \\
NGC 3256\tablenotemark{e} & $>13$ & $>41.0$ & $44.9$ & $17.4$ & $>0.7$ \\  
M101\tablenotemark{f} &  $>5$ & $>39.6$ & $43.6$ &  $0.9$ & $>5.6$ \\
\enddata


\tablenotetext{a}{Number of sources with $L_{\rm x} > 2.1 \times
10^{38}$~erg~s$^{-1}$ in the $0.4$--$8$~keV band.}
\tablenotetext{b}{Total luminosity from sources with $L_{\rm x} >
2.1 \times 10^{38}$~erg~s$^{-1}$ in the $0.4$--$8$~keV band.}
\tablenotetext{c}{Observed $42.5$--$122.5\mu$m luminosity (see
\S~\ref{sec-galaxies}).}
\tablenotetext{d}{Star formation rate (see \S~\ref{sec-galaxies}).}
\tablenotetext{e}{The limiting luminosity level during the
\emph{Chandra} observation was $\simeq 10^{39}$ erg s$^{-1}$ in the
central region of this galaxy.}
\tablenotetext{f}{The galactic disk of M101 at the 25.0 B-magnitude
(arc sec)$^{-2}$ isophote level subtends a much larger solid angle
than does the S3 chip.}

\end{deluxetable}


\begin{thebibliography}{}
\bibitem[Abramowicz et al.(1980)]{abr80} Abramowicz, M. A., Calvani,
    M., \& Nobili, L. 1980, ApJ, 242, 772
\bibitem[Allen et al.(1978)]{all78} Allen, R. J., van der Hulst,
    J. M., Goss, W. M., \& Huchtmeier, W. 1978, \aap, 64, 359
\bibitem[Anders \& Grevesse(1989)]{ag89} Anders, E., \& Grevesse,
    N. 1989, Geochimica et Cosmochimica Acta, 53, 197
\bibitem[Arnaud(1996)]{arn96} Arnaud, K. A. 1996, in ASP Conf. Proc. 101,
    Astronomical Data Analysis Software and Systems V, ed. G. Jacoby \&
    J. Barnes (San Francisco: ASP), 17
\bibitem[Bauer et al.(2001)]{bau01} Bauer, F. E., Brandt, W. N.,
    Sambruna, R. M., Chartas, G., Garmire, G. P., Kaspi, S., \& Netzer,
    H. 2001, \aj, 122, 182
\bibitem[Begelman(2002)]{beg02} Begelman, M. C. 2002, \apj, 568, L97
\bibitem[Bevington \& Robinson(1992)]{br92} Bevington, P. R., \&
    Robinson, D. K. 1992, Data Reduction and Error Analysis for the
    Physical Sciences (2nd ed.; New York: McGraw-Hill)
\bibitem[Bland-Hawthorn et al.(1997)]{bla97} Bland-Hawthorn, J.,
    Gallimore, J. F., Tacconi, L. J., Brinks, E., Baum, S. A., Antonucci,
    R. R. J., \& Cecil, G. N. 1997, \apss, 248, 9
\bibitem[Blanton et al.(2001)]{bla01} Blanton, E. L., Sarazin, C. L.,
    \& Irwin, J. A. 2001, \apj, 552, 106
\bibitem[Bond et al.(1984)]{bon84} Bond, J. R., Arnett, W. D., \&
    Carr, B. J. 1984, \apj, 280, 825
\bibitem[Colbert \& Mushotzky(1999)]{cm99} Colbert, E. J. M., \&
    Mushotzky, R. F. 1999, \apj, 519, 89
\bibitem[Colbert \& Ptak(2002)]{cp02} Colbert, E. J. M., \& Ptak,
    A. F. 2002, \apjs, 143, 25
\bibitem[Condon(1992)]{con92} Condon, J. J. 1992, \araa, 30, 575
\bibitem[Cowie et al.(2002)]{cow02} Cowie, L. L., Garmire, G. P.,
    Bautz, M. A., Barger, A. J., Brandt, W. N., \& Hornschemeier,
    A. E. 2002, \apj, 566, L5
\bibitem[de Vaucouleurs et al.(1991)]{dev91} de Vaucouleurs, G., de
    Vaucouleurs, A., Corwin, H. G., Buta, R. J., Paturel, G., \& Fouqu\'{e},
    P. 1991, Third Reference Catalogue of Bright Galaxies (New York:
    Springer)
\bibitem[Devereux \& Young(1990)]{dy90} Devereux, N. A., \& Young,
    J. S. 1990, \apj, 350, L25
\bibitem[Dickey \& Lockman(1990)]{dl90} Dickey, J. M., \& Lockman,
    F. J. 1990, \araa, 28, 215
\bibitem[Dotani et al.(1997)]{dot97} Dotani, T., et al. 1997, \apj,
    485, L87
\bibitem[Ebisawa et al.(1994)]{ebi94} Ebisawa, K., et al. 1994, \pasj,
    46, 375
\bibitem[Ebisuzaki et al.(2001)]{ebi01} Ebisuzaki, T., et al. 2001,
    \apj, 562, L19
\bibitem[Esin et al.(1997)]{esi97} Esin, A. A., McClintock, J. E., \&
    Narayan, R. 1997, \apj, 489, 865
\bibitem[Fabbiano(1995)]{fab95} Fabbiano, G. 1995, in X-ray Binaries,
    ed. W. H. G. Lewin, J. van Paradijs, \& E. P. J. van den Heuvel
    (Cambridge: Cambridge Univ. Press), 390
\bibitem[Fabian \& Terlevich(1996)]{ft96} Fabian, A. C., \& Terlevich,
    R. 1996, \mnras, 280, L5
\bibitem[Feldmeier et al.(1997)]{fel97} Feldmeier, J. J., Ciardullo,
    R., \& Jacoby, G. H. 1997, \apj, 479, 231
\bibitem[Fox et al.(2000)]{fox00} Fox, D. W., et al. 2000, \mnras,
    319, 1154
\bibitem[Franco et al.(1993)]{fra93} Franco, J., Miller, W., Cox, D.,
    Terlevich, R., R\'{o}\.{z}yczka, M., \& Tenorio-Tagle, G. 1993,
    Rev. Mexicana Astron. Astrofis., 27, 133
\bibitem[Freeman et al.(1977)]{fre77} Freeman, K. C., Karlsson, B.,
    Lyng\aa, G., Burrell, J. F., van Woerden, H., \& Goss, W. M. 1977,
    \aap, 55, 445
\bibitem[Freeman et al.(2002)]{fre02} Freeman, P. E., Kashyap, V.,
    Rosner, R., \& Lamb, D. Q. 2002, \apjs, 138, 185
\bibitem[Fryer(1999)]{fry99} Fryer, C. L. 1999, \apj, 522, 413
\bibitem[Fryer \& Kalogera(2001)]{fk01} Fryer, C. L., \& Kalogera, V.,
    2001, \apj, 554, 548
\bibitem[Fryer et al.(2001)]{fry01} Fryer, C. L., Woosley, S. E., \&
    Heger, A., 2001, \apj, 550, 372
\bibitem[Garmire(1997)]{gar97} Garmire, G. P. 1997, \baas, 190, 34.04
\bibitem[Gehrels(1986)]{geh86} Gehrels, N. 1986, \apj, 303, 336
\bibitem[Glatzel et al.(1985)]{gla85} Glatzel, W., El Eid, M. F., \&
    Fricke, K. J. 1985, \aap, 149, 413
\bibitem[Grimm et al.(2002)]{gri02} Grimm, H.-J., Gilfanov, M., \&
    Sunyaev, R. 2002, \aap, 391, 923
\bibitem[Grimm et al.(2003)]{gri03} Grimm, H.-J., Gilfanov, M., \&
    Sunyaev, R. 2003, \mnras, 339, 793
\bibitem[Kaaret et al.(2001)]{kaa01} Kaaret, P., Prestwich, A. H.,
    Zezas, A., Murray, S. S., Kim, D.-W.; Kilgard, R. E., Schlegel, E. M.,
    \& Ward, M. J. 2001, \mnras, 321, L29
\bibitem[Kilgard et al.(2002)]{kil02} Kilgard, R. E., Kaaret, P.,
    Krauss, M. I., Prestwich, A. H., Raley, M. T., \& Zezas, A. 2002,
    \apj, 573, 138
\bibitem[Kim \& Fabbiano(2002)]{kf02} Kim, D.-W., \& Fabbiano, G. 2002,
    \apj, submitted (astro-ph/0206369)
\bibitem[King et al.(2001)]{kin01} King, A. R., Davies, M. B., Ward,
    M. J., Fabbiano, G., \& Elvis, M. 2001, \apj, 552, L109
\bibitem[Klein et al.(1996)]{kle96} Klein, R. I., Arons, J., Jernigan,
    G., \& Hsu, J. J.-L. 1996, \apj, 457, L85
\bibitem[K\"{o}rding et al.(2002)]{kor02} K\"{o}rding, E., Falcke, H.,
    \& Markoff, S. 2002, \aap, 382, L13
\bibitem[Kraft et al.(1991)]{kra91} Kraft, R. P., Burrows, D. N., \&
    Nousek, J. A. 1991, \apj, 374, 344
\bibitem[Kubota et al.(1998)]{kub98} Kubota, A., Tanaka, Y.,
    Makishima, K., Ueda, Y., Dotani, T., Inoue, H., \& Yamaoka, K. 1998,
    \pasj, 50, 667
\bibitem[Kubota et al.(2001)]{kub01} Kubota, A., Mizuno, T.,
    Makishima, K., Fukazawa, Y., Kotoku, J., Ohnishi, T., \& Tashiro,
    M. 2001, \apj, 547, L119
\bibitem[Kubota et al.(2003)]{kub03} Kubota, A., Done, C., \&
    Makishima, K. 2003, \mnras, 337, L11
\bibitem[La Parola et al.(2001)]{lap01} La Parola, V., Peres, G.,
    Fabbiano, G., Kim, D. W., \& Bocchino, F. 2001, \apj, 556, 47
\bibitem[L\'ipari et al.(2000)]{lip00} L\'ipari, S., D\'iaz, R.,
    Taniguchi, Y., Terlevitch, R., Dottori, H., \& Carranza, G. 2000, \aj,
    120, 645
\bibitem[Lira et al.(2000)]{lir00} Lira, P., Lawrence, A., \& Johnson,
    R. J.  2000, \mnras, 319, 17
\bibitem[Lira et al.(2002)]{lir02} Lira, P., Ward, M., Zezas, A.,
    Alonso-Herrero, A., \& Ueno, S. 2002, \mnras, 330, 259
\bibitem[Liu et al.(2000)]{liu00} Liu, Q. Z., van Paradijs, J., \& van
    den Heuvel, E. P. J. 2001, \aaps, 147, 25
\bibitem[Madau \& Rees(2001)]{mr01} Madau, P., \& Rees, M. J. 2001,
    \apj, 551, L27
\bibitem[Makishima et al.(1986)]{mak86} Makishima, K., Maejima, Y.,
    Mitsuda, K., Bradt, H. V., Remillard, R. A., Tuohy, I. R., Hoshi, R.,
    \& Nakagawa, M. 1986, \apj, 308, 635
\bibitem[Makishima et al.(2000)]{mak00} Makishima, K., et al. 2000,
    \apj, 535, 632
\bibitem[Markoff et al.(2001)]{mar01} Markoff, S., Falcke, H., \&
    Fender, R. 2001, \aap, 372, L25
\bibitem[Matsumoto \& Tsuru(1999)]{mt99} Matsumoto, H., \& Tsuru,
    T. G. 1999, \pasj, 51, 321
\bibitem[Matsumoto et al.(2001)]{mat01} Matsumoto, H., Tsuru, T. G.,
    Koyama, K., Awaki, H., Canizares, C. R., Kawai, N., Matsushita, S., \&
    Kawabe, R. 2001, \apj, 547, L25
\bibitem[Miller(1996)]{mil96} Miller, G. S. 1996, \apj, 468, L29
\bibitem[Miller \& Hamilton(2002)]{mh02} Miller, M. C., \& Hamilton,
    D. P. 2002, \mnras, 330, 232
\bibitem[Mitsuda et al.(1984)]{mit84} Mitsuda, K., et al. 1984, \pasj,
    36, 741
\bibitem[Mizuno et al.(1999)]{miz99} Mizuno, T., Ohnishi, T., Kubota,
    A., Makishima, K., \& Tashiro, M. 1999, \pasj, 51, 663
\bibitem[Mizuno et al.(2001)]{miz01} Mizuno, T., Kubota, A., \&
    Makishima, K. 2001, \apj, 554, 1282
\bibitem[Morrison \& McCammon(1983)]{mm83} Morrison, R., \& McCammon,
    D. 1983, \apj, 270, 119
\bibitem[Moshir et al.(1990)]{mos90} Moshir, M., et al. 1990, 
    \emph{IRAS} Faint Source Catalog, Ver. 2.0 (Greenbelt: NASA/GSFC)
\bibitem[Mukai(1993)]{muk93} Mukai, K. 1993, Legacy 3, 21
\bibitem[Nowak(1995)]{now95} Nowak, M. A. 1995, \pasp, 107, 1207
\bibitem[Okada et al.(1998)]{oka98} Okada, K., Dotani, T., Makishima,
    K., Mitsuda, K., \& Mihara, T. 1998, \pasj, 50, 25
\bibitem[Pence et al.(2001)]{pen01} Pence, W. D., Snowden, S. L.,
    Mukai, K., \& Kuntz, K. D. 2001, \apj, 561, 189
\bibitem[Plewa(1995)]{ple95} Plewa, T. 1995, \mnras, 275, 143
\bibitem[Pogge \& de Robertis(1993)]{pd93} Pogge, R. W., \& De
    Robertis, M. M. 1993, \apj, 404, 563
\bibitem[Portegies Zwart \& McMillan(2002)]{pm02} Portegies Zwart,
    S. F., \& McMillan, S. L. W. 2002, \apj, 576, 899
\bibitem[Press et al.(1992)]{pre92} Press, W. H., Teukolsky, S. A.,
    Vetterling, W. T., \& Flannery, B. P. 1992, Numerical Recipes in C
    (2nd ed.; Cambridge: Cambridge Univ. Press)
\bibitem[Reynolds et al.(1997)]{rey97} Reynolds, C. S., Loan, A. J.,
    Fabian, A. C., Makishima, K., Brandt, W. N., \& Mizuno, T. 1997,
    \mnras, 286, 349
\bibitem[Rice et al.(1988)]{ric88} Rice, W., et al. 1988, \apjs, 68, 91
\bibitem[Roberts \& Warwick(2000)]{rw00} Roberts, T. R., \& Warwick,
    R. S. 2000, \mnras, 315, 98
\bibitem[Sakai \& Madore(1999)]{sm99} Sakai, S., \& Madore,
    B. F. 1999, \apj, 526, 599
\bibitem[Sandage \& Bedke(1994)]{sb94} Sandage, A., \& Bedke, J. 1994,
    The Carnegie Atlas of Galaxies (Washington, DC: Carnegie Institution
    of Washington with The Flintridge Foundation)
\bibitem[Sarazin et al.(2000)]{sar00} Sarazin, C. L., Irwin, J. A., \&
    Bregman, J. N. 2000, \apj, 544, L101
\bibitem[Shimura \& Takahara(1995)]{st95} Shimura, T., \&
    Takahara, F. 1995, \apj, 445, 780
\bibitem[Smith \& Wilson(2001)]{sw01} Smith, D. A., \& Wilson,
    A. S. 2001, \apj, 557, 180
\bibitem[Stetson et al.(1998)]{ste98} Stetson, P. B., et al. 1998,
    \apj, 508, 491
\bibitem[Strickland et al.(2001)]{str01} Strickland, D. K., Colbert,
    E. J. M., Heckman, T. M., Weaver, K. A., Dahlem, M., \& Stevens,
    I. R. 2001, \apj, 560, 707
\bibitem[Sugiho et al.(2001)]{sug01} Sugiho, M., Kotoku, J.;
    Makishima, K., Kubota, A., Mizuno, T., Fukazawa, Y., \& Tashiro,
    M. 2001, \apj, 561, L73
\bibitem[Tanaka(2000)]{tan00a} Tanaka, Y. 2000, in IAU Symp. 195,
    Highly Energetic Physical Processes and Mechanisms for Emission from
    Astrophysical Plasmas, eds. P. C. H. Martens, S. Tsuruta, \&
    M. A. Weber (San Francisco: ASP), 37
\bibitem[Taniguchi et al.(2000)]{tan00b} Taniguchi, Y., Shioya, Y.,
    Tsuru, T. G., \& Ikeuchi, S. 2000, \pasj, 52, 533
\bibitem[Telesco \& Decher(1988)]{td88} Telesco, C. M., \& Decher,
    R. 1988, \apj, 334, 573
\bibitem[Terashima \& Wilson(2002a)]{tw02a} Terashima, Y., \& Wilson,
    A. S. 2002a, ApJ, in preparation
\bibitem[Terashima \& Wilson(2002b)]{tw02b} Terashima, Y., \& Wilson,
    A. S. 2002b, in Proceedings of the Symposium on ``New Visions of the
    Universe in the XMM-Newton and Chandra Era,'' ed. F. Jansen (European
    Space Agency ESA SP-488), in press (astro-ph/0204321)
\bibitem[Tremaine et al.(1975)]{tre75} Tremaine, S. D., Ostriker,
    J. P., \& Spitzer L., Jr. 1975, \apj, 196, 407
\bibitem[Ueda et al.(1999)]{ued99} Ueda, Y., Takahashi, T., Ishisaki,
    Y., Ohashi, T., \& Makishima, K. 1999, \apj, 524, L11
\bibitem[van Paradijs \& McClintock(1995)]{vm95} van Paradijs, J., \&
    McClintock, J. E. 1995, in X-ray Binaries, ed. W. H. G. Lewin, J. van
    Paradijs, \& E. P. J. van den Heuvel (Cambridge: Cambridge
    Univ. Press), 58
\bibitem[van Speybroeck(1997)]{van97} van Speybroeck, L. P. 1997,
    \baas, 190, 34.03
\bibitem[Vink et al.(2001)]{vin01} Vink, J. S., de Koter, A., \&
    Lamers, H. J. G. L. M. 2001, \aap, 369, 574
\bibitem[Watarai et al.(2000)]{wat00} Watarai, K., Fukue, J.,
    Takeuchi, M., \& Mineshige, S. 2000, \pasj, 52, 133
\bibitem[Watarai et al.(2001)]{wat01} Watarai, K., Mizuno, T., \&
    Mineshige, S. 2001, \apj, 549, L77
\bibitem[Watarai(2002)]{wat02} Watarai, K. 2002, private communication
\bibitem[Weedman(1983)]{wee83} Weedman, D. W. 1983, \apj, 266, 479
\bibitem[Weisskopf et al.(2001)]{wei01} Weisskopf, M. C., Brinkman,
    B., Canizares, C., Garmire, G., Murray, S., \& Van Speybroeck,
    L. P. 2001, \pasp, 114, 1
\bibitem[White et al.(1995)]{whi95} White, N. E., Nagase, F., \&
    Parmar, A. N. 1995, in X-ray Binaries, ed. W. H. G. Lewin, J. van
    Paradijs, \& E. P. J. van den Heuvel (Cambridge: Cambridge
    Univ. Press), 1
\bibitem[Woosley(1986)]{woo86} Woosley, S. E. 1986, in Nucleosynthesis
    and Chemical Evolution, 16th Adv. Course, Saas-Fe, ed. B. Hauck,
    A. Maeder, \& G. Meynet (CH-1290 Sauverny-Versoix: Geneva
    Observatory), 1
\bibitem[Woosley \& Weaver(1982)]{ww82} Woosley, S. E., \& Weaver,
    T. A. 1982, in Supernovae: A Survey of Current Research,
    ed. M. J. Rees \& R.  J. Stoneham (Dordrecht: Reidel), 79
\bibitem[Young et al.(2001)]{you01} Young, A. J., Wilson, A. S., \&
    Shopbell, P. L. 2001, \apj, 556, 6 (Paper~I)
\bibitem[Zezas et al.(1999)]{zez99} Zezas, A. L., Georgantopoulos, I.,
    \& Ward, M. J. 1999, \mnras, 308, 302
\bibitem[Zezas et al.(2002)]{zez02} Zezas, A., Fabbiano, G., Rots,
    A. H., \& Murray, S. S. 2002, \apjs, 142, 239
\bibitem[Zezas \& Fabbiano(2002)]{zf02} Zezas, A., \& Fabbiano, G.
    2002, \apj, 577, 726
\bibitem[Zhang et al.(1997)]{zha97} Zhang, S. N., Cui, W., \& Chen,
    W. 1997, \apj, 482, L155
\bibitem[Zycki et al.(1999)]{zyc99} \.{Z}ycki, P. T., Done, C., \&
    Smith, D. A. 1999, \mnras, 309, 561
\end{thebibliography}
\end{document}